\documentclass[preprint,showpacs,amsmath,amssymb,aps,prd,nofootinbib]{revtex4}
\usepackage{epsfig,graphicx,color,appendix}
\begin{document}

\title{\mbox{}\\[10pt]
Quark-lepton complementarity and tribimaximal neutrino mixing from discrete symmetry }

\author{Y. H. Ahn\footnote{Email: yhahn@phys.sinica.edu.tw},
Hai-Yang Cheng\footnote{Email: phcheng@phys.sinica.edu.tw},
and Sechul Oh\footnote{Email: scoh@phys.sinica.edu.tw}}

\affiliation{Institute of Physics, Academia Sinica, Taipei 115, Taiwan}

\date{\today}

\begin{abstract}
The quark-lepton complementarity (QLC) relations indicate a deep structure that interrelates quarks
and leptons.  We propose new scenarios, in a seesaw framework with discrete $A_4$ flavor symmetry,
which can accommodate the QLC relations and the nonzero neutrino mixing angle $\theta_{13}$ together
with all the available neutrino experimental data, in a consistent way to generate the
Cabibbo-Kobayashi-Maskawa (CKM) matrix for the quark mixing.
Certain effective dimension-5 operators are introduced, which induce a deviation of the lepton mixing
matrix from the tribimaximal mixing (TBM) pattern and lead the quark mixing matrix to the CKM one
in form.
We explicitly demonstrate three different possibilities of constructing the charged lepton mixing
matrix and point out that the {\it phases} of whose elements play a crucial role to satisfy the QLC
relations. We find that for the reactor mixing angle $\theta_{13}$ its possible values can vary around
the center value $\sin \theta_{13} \simeq \lambda /\sqrt{2}$ ($\lambda \simeq 0.22$ being the Cabbibo
angle) and have the lower bound $\theta_{13} \gtrsim 3.5^{\circ}$.  We also show that sizable leptonic
CP violation characterized by the Jarlskog invariant $|J_{\rm CP}|\sim{\cal O}(10^{-2})$ is allowed,
which is expected to be tested in the future experiments such as the upcoming long baseline neutrino
oscillation ones.
\end{abstract}


\maketitle
\section{Introduction}
Recent analyses on the knowledge of neutrino oscillation parameters, which enter into a new phase of
the precise determination of mixing angles and mass squared differences as shown in Table~\ref{tab:data},
indicate that neutrinos are massive and leptons of different flavors mix with each other through the
charged weak interactions.
\begin{table}[th]
\begin{center}
\caption{\label{tab:data} Current best-fit values as well as $1\sigma$ and $3\sigma$ ranges of the
oscillation parameters~\cite{GonzalezGarcia:2010er}.  Here ``Nor" and ``Inv" indicate the
normal and inverted mass orderings of neutrinos, respectively.}
\begin{tabular}{|c|c|c|c|c|c|} \hline
 & $\Delta m^{2}_{\rm sol}/10^{-5}\mathrm{\ eV}^2$ & $\theta_{12}$ & $\theta_{13}$ & $\theta_{23}$ &
$\Delta m^{2}_{\rm atm}/10^{-3}\mathrm{\ eV}^2$  Nor(Inv) \\ \hline \hline
Best-fit  &     7.59     &  $34.4^{\circ}$  &  $5.6^{\circ}$  & $42.8^{\circ}$  &  $2.46~~~(-2.36)$
\\ \hline
$1\sigma$ & $7.39 - 7.79$ & $33.4^{\circ}-35.4^{\circ}$  & $2.9^{\circ}-8.6^{\circ}$
 & $39.9^{\circ}-47.5^{\circ}$  & $2.46\pm0.12~~(-2.36\pm0.11)$ \\ \hline
$3\sigma$ & $6.90 - 8.20$ & $31.5^{\circ} - 37.6^{\circ}$  & $<12.5^{\circ}$
 & $35.5^{\circ}-53.5^{\circ}$  & $2.46\pm0.37~~(-2.36\pm0.37)$ \\ \hline
\end{tabular}
\end{center}
\end{table}
At present, the experimental data at $3\sigma$ level~\cite{GonzalezGarcia:2010er,nudata}
are fully consistent with the tribimaximal mixing (TBM) pattern suggested by Harrison,
Perkins and Scott~\cite{Harrison:2002er} with the mixing angles
 \begin{eqnarray}
  \sin^{2}\theta_{12}=\frac{1}{3},~~~~~~\sin^{2}\theta_{23}=\frac{1}{2},~~~~~~\sin\theta_{13}=0~.
  \label{tibi1}
 \end{eqnarray}
However, the recent analyses based on global fits of the available data give us hints for $\theta_{13}>0$
at $1\sigma$ level~\cite{GonzalezGarcia:2010er,nudata}, which requires some deviation of the mixing
angles from the TBM pattern.
Although neutrinos have gradually revealed their properties in various experiments since the historic
Super-Kamiokande confirmation of neutrino oscillations~\cite{Fukuda:1998mi}, properties related to the
leptonic CP violation are completely unknown yet.  In addition, the {\it large} values of
the solar mixing angle $\theta_{\rm sol}\equiv\theta_{12}$ and the atmospheric mixing angle
$\theta_{\rm atm}\equiv\theta_{23}$ may be telling us about some new flavor symmetries of leptons absent
in the quark sector with the {\it small} quark mixing angles, e.g., at $1\sigma$
level~\cite{ckmfitter}
 \begin{eqnarray}
  \theta^{q}_{12} = ( 13.03 \pm 0.05 )^{\circ}~,
  ~~\theta^{q}_{23} = ( 2.37^{+0.05}_{-0.09} )^{\circ} ~
  ~~\theta^{q}_{13} = ( 0.20^{+0.02}_{-0.02} )^{\circ} ~,
  ~~\delta^{q}_{CP} = ( 67.17_{-2.44}^{+2.78} )^{\circ} ~.
 \end{eqnarray}
This fact may provide a clue to the nature of quark-lepton physics beyond the Standard Model (SM).
Therefore, it is very important to find a natural model that leads to the observed flavor mixing
patterns for quarks and leptons with good accuracy. The disparity that nature exhibits between
the quark and lepton mixing angles has been suggested in terms of the quark-lepton complementarity
(QLC) relations~\cite{Raidal:2004iw,Minakata:2004xt,QLC},
which can be written as
 \begin{eqnarray}
  \theta_{12}+\theta^{q}_{12}\simeq45^{\circ}~,~~~\theta_{23}+\theta^{q}_{23}\simeq45^{\circ}~.
  \label{QLC}
 \end{eqnarray}
The QLC relations indicate that there could be a quark-lepton symmetry based on a flavor symmetry.
In the past few years there have been lots of efforts in searching for models which can reproduce
the TBM pattern for the neutrino mixing matrix.  A fascinating way seems to be the use of
certain discrete non-Abelian flavor groups added to the gauge groups of the SM.
The $\mu-\tau$ symmetry, which is the most popular discrete symmetry, has made some success
in describing the mass and mixing patterns in the leptonic sector~\cite{mutau}. Further, Ma and
Rajasekaran~\cite{Ma:2001dn} have introduced for the first time the $A_{4}$ flavor symmetry to avoid
the mass degeneracy between $\mu$ and $\tau$ under the $\mu-\tau$ symmetry. Many subsequent works in
the literature invoke the same symmetry group $A_{4}$.  One reason is that $A_{4}$ is rather economical:
it is the smallest discrete group containing a three-dimensional irreducible representation.
In well-motivated extensions~\cite{He:2006dk,Altarelli:2005yp} of the SM through the inclusion of the
$A_{4}$ discrete symmetry, the TBM pattern for the neutrino mixing matrix comes out in a natural way.
Especially, the authors of Ref.~\cite{He:2006dk} have suggested dimension-5 operators induced in the
neutrino sector as a natural source for the quark mixing in an economical way.

In this work we present new scenarios, based on the $A_4$ flavor symmetry, that can
accommodate the QLC relations and the nonzero mixing angle $\theta_{13}$ together with all the other
neutrino experimental data, such as $\Delta m_{\rm sol}^2$, $\Delta m_{\rm atm}^2$, $\sin^2 \theta_{12}$
and $\sin^2 \theta_{23}$, in the same framework to generate the Cabibbo-Kobayashi-Maskawa (CKM) matrix
for the quark mixing.
We extend the framework of Ref.~\cite{He:2006dk} by introducing all possible effective dimension-5
operators, invariant under $SU(2)_{\rm L} \times U(1)_{\rm Y} \times A_{4} \times Z_{2}$ symmetry in
a seesaw framework\footnote{In our scenarios, after $A_4$ symmetry breaking, there are
no residual symmetries like $Z_{2}$ and $C_{3}$, neither in the neutrino nor in the quark
sector, on the contrary to that of Ref.~\cite{He:2006dk}.},
{\it both} in the neutrino sector {\it and} in the charged fermion (quarks and charged leptons)
sector\footnote{These dimension-5 operators can induce a source of the high energy
CP violation responsible for leptogenesis~\cite{review}, which we are not going to study in this work.}.
Due to the higher dimensional operators introduced in the neutrino and charged lepton sectors, the
lepton mixing matrix has a deviation from the TBM pattern which can explain the nonzero $\theta_{13}$,
while the higher dimensional operators appearing in the quark sector lead the quark mixing matrix to
the CKM one in form.  Thus, the large lepton mixing and the small quark mixing observed by experiments
are understood in a natural way.

In our framework base on the $A_4$ flavor symmetry, the charged lepton and quark sectors
have the same flavor structure in the Lagrangian, implying that their mixing structures could also be
related to each other.
Motivated by this point, we further explore a possibility that the QLC relations can be understood by
certain underlying relations between the charged lepton and quark mixings.
Starting from the fact that the mixing matrix of the up-type quark sector can be almost diagonal and
so the CKM matrix is mainly generated from the down-type quark mixing matrix, we assume that the
mixing matrix of the charged lepton sector is basically the same in form as that of the down-type quark
sector, except for the possibly different {\it phases} of each matrix element.
Consequently, the lepton mixing matrix appears as the multiplication of the ``CKM-like matrix''
(induced from the charged lepton sector) and the ``TBM pattern matrix'' (induced from the neutrino
sector), where each element of both matrices has an arbitrary {\it phase}.
In our framework, the dimension-5 operators generate all the necessary off-diagonal elements of each
mixing matrix induced, respectively, from the neutrino, charged lepton and quark sectors.
We then show that the QLC relations can be satisfied by the relevant elements of the lepton mixing
matrix in the above particular form.  It turns out that certain {\it phases} of the CKM-like matrix
elements induced from the {\it charged lepton sector} plays a crucial role to satisfy the QLC relations.
It should be emphasized that this feature is very different from the conventional QLC scenario
which is characterized by the ``bimaximal minus CKM mixing''~\cite{Minakata:2004xt,qlc}.
For our aim, we demonstrate in detail three possible scenarios corresponding to three different
ways of constructing the charged lepton mixing matrix, led by different assumptions on the
hierarchical charged fermion Yukawa couplings of the relevant dimension-5 operators.
We will see that all these three scenarios can lead to the QLC relations. Finally we shall elaborate on
the other phenomenological consequences, including
the lower
bound on the mixing angle $\theta_{13}$ and the sizable leptonic CP violation characterized
by the Jarlskog invariant $|J_{\rm CP}|$.

The paper is organized as follows. In Sec.~II, we present the particle content together with the flavor
symmetry of our model. Then, after introducing the effective Lagrangians for the neutrino and charged
fermion sectors, we derive the realistic mixing matrices for each sector.
In Sec.~III, we show, both analytically and numerically, how the QLC relations can be realized by the
relevant elements of the lepton mixing matrix. Other phenomenological consequences including the
nonzero $\theta_{13}$ and the leptonic CP violation are also discussed. Finally, our conclusions are
summarized in Sec.~IV.  In Appendix, we show the Higgs potential and the relevant vacuum alignment.

\section{The framework with $A_4 \times Z_2$ symmetry}

We work in the framework of an extension of the SM by introducing the extra right-handed
$SU(2)_{L}$-singlet Majorana neutrinos $N_{R}$. Unless flavor symmetries are assumed, particle masses
and mixings are generally undetermined in gauge theory.
To understand the present neutrino oscillation data and the quark mixing data, we consider the $A_{4}$
flavor symmetry together with an auxiliary symmetry $Z_2$ for leptons and quarks.  Then the symmetry
group for the lepton and quark sectors is ~$SU(2)_L \times U(1)_Y \times A_{4} \times Z_{2}$.~
To impose the $A_4$ flavor symmetry on our model properly, apart from the usual SM Higgs doublet $\Phi$,
the scalar sector is extended by introducing two types of new scalar fields, $\chi$ and $\eta$, that
are an $SU(2)_L$-singlet and an $SU(2)_L$-doublet, respectively:
 \begin{eqnarray}
  \Phi =
  {\left(\begin{array}{c}
  \varphi^{+} \\
  \varphi^{0}
 \end{array}\right)}~,~~~\chi~,~~~\eta =
  {\left(\begin{array}{c}
  \eta^{+} \\
  \eta^{0}
 \end{array}\right)}~.
  \label{Higgs}
 \end{eqnarray}
Here we recall that $A_{4}$ is the symmetry group of a tetrahedron and the finite group of the even
permutation of four objects.
Its irreducible representations contain one triplet ${\bf 3}$ and three singlets ${\bf 1}$, ${\bf 1}'$,
${\bf 1}''$ with the multiplication rules
${\bf 3}\otimes{\bf 3} = {\bf 3}_{s}\oplus{\bf 3}_{a}\oplus{\bf 1}\oplus{\bf 1}'\oplus{\bf 1}''$,~
${\bf 1}'\otimes{\bf 1}'' = {\bf 1}$,~ ${\bf 1}'\otimes{\bf 1}' = {\bf 1}''$~ and~
${\bf 1}''\otimes{\bf 1}'' = {\bf 1}'$.
By denoting $(a_{1}, a_{2}, a_{3})$ and $(b_{1}, b_{2}, b_{3})$ as two $A_4$ triplets, we obtain
 \begin{eqnarray}
  (a\otimes b)_{{\bf 3}_{\rm s}} &=& (a_{2}b_{3}+a_{3}b_{2}, a_{3}b_{1}+a_{1}b_{3},
   a_{1}b_{2}+a_{2}b_{1})~, \nonumber\\
  (a\otimes b)_{{\bf 3}_{\rm a}} &=& (a_{2}b_{3}-a_{3}b_{2}, a_{3}b_{1}-a_{1}b_{3},
   a_{1}b_{2}-a_{2}b_{1})~, \nonumber\\
  (a\otimes b)_{{\bf 1}} &=& a_{1}b_{1}+a_{2}b_{2}+a_{3}b_{3}~,\nonumber\\
  (a\otimes b)_{{\bf 1}'} &=& a_{1}b_{1}+\omega a_{2}b_{2}+\omega^{2}a_{3}b_{3}~,\nonumber\\
  (a\otimes b)_{{\bf 1}''} &=& a_{1}b_{1}+\omega^{2} a_{2}b_{2}+\omega a_{3}b_{3}~,
 \end{eqnarray}
where $\omega=e^{i2\pi/3}$ is a complex cubic-root of unity.

The field content under $SU(2)_L \times U(1)_Y \times A_{4} \times Z_{2}$ in our model is assigned in
Table~\ref{reps}, where $L_{L}=(\nu_{L},\ell^{-}_{L})^T$ and $Q_{L}=(u_{L},d_{L})^T$ are the SM
left-handed lepton and quark doublets, respectively, and $l_R$ and $u_R ~(d_R)$ are the respective SM
right-handed lepton and u-type (d-type) quark singlets.
In our framework, we assume that there is a cutoff scale $\Lambda$, above which there exists unknown
physics with no CP violation term. Then below the scale $\Lambda$, the higher dimensional operators
express the effects from the unknown physics.

\begin{table}[h]
\caption{\label{reps}
Representations of the fields under $SU(2)_L \times U(1)_Y \times A_{4} \times Z_{2}$.}
\begin{ruledtabular}
\begin{tabular}{ccccccccccccc}
Field &$L_{L}$&$Q_{L}$&$l_R,l'_R,l''_R$&$u_R,u'_R,u''_R$&$d_R,d'_R,d''_R$&$N_{R}$&$\chi$
 &$\Phi$&$\eta$
\\ \hline
$A_4$&$\mathbf{3}$&$\mathbf{3}$&$\mathbf{1}$, $\mathbf{1^\prime}$,$\mathbf{1^{\prime\prime}}$
 &$\mathbf{1}$, $\mathbf{1^\prime}$,$\mathbf{1^{\prime\prime}}$
 &$\mathbf{1}$, $\mathbf{1^\prime}$,$\mathbf{1^{\prime\prime}}$
 &$\mathbf{3}$ &$\mathbf{3}$&$\mathbf{3}$&$\mathbf{1}$\\
$Z_2$&$+$&$+$&$+$&$+$&$+$&$-$&$+$&$+$&$-$\\
$SU(2)_L\times U(1)_Y$&$(2,-1)$&$(2,\frac{1}{3})$&$(1,-2)$&$(1,\frac{4}{3})$
 &$(1,-\frac{2}{3})$&$(1,0)$&$(1,0)$&$(2,1)$&$(2,1)$\\
\end{tabular}
\end{ruledtabular}
\end{table}

\subsection{The neutrino sector}

With dimension-5 operators driven by the $\chi$ field, the Yukawa interactions $(d\leq5)$ in the
neutrino sector, invariant under $SU(2)\times U(1)\times A_{4}\times Z_{2}$, can be written as
 \begin{eqnarray}
 -{\cal L}^{\nu}_{\rm Yuk} &=& y_{\nu}(\bar{L}_{L}N_{R})_{{\bf 1}}\tilde{\eta}
  +\frac{1}{2}M [ (\overline{N_R})^c N_R ]_{{\bf 1}}
  +\frac{1}{2}\lambda_{\chi}^s [ (\overline{N_R})^c N_R ]_{{\bf 3}_{s}} \chi\nonumber\\
 &+&\frac{y^{s}_{N}}{\Lambda}[(\bar{L}_{L}N_{R})_{{\bf 3}_{s}}\chi]\tilde{\eta}
  +\frac{y^{a}_{N}}{\Lambda}[(\bar{L}_{L}N_{R})_{{\bf 3}_{a}}\chi]\tilde{\eta}+ {\rm H.c.} ~,
 \label{lagrangian}
 \end{eqnarray}
where $\tilde{\eta}\equiv i\tau_{2}\eta^{\ast}$ and $\tau_{2}$ is the Pauli matrix.
In the above Lagrangian, the right-handed Majorana neutrino terms are associated with a bare mass $M$
and an SM gauge singlet scalar field $\chi$ which is a $A_4$ triplet. There is no ${\bf 3}_a$ Majorana
neutrino term, since the term $[ (\overline{N_R})^c N_R ]_{{\bf 3}_a}$ identically vanishes due to the
property of a Majorana particle. By imposing  the additional symmetry $Z_{2}$ as shown in
Table~\ref{reps}, the $SU(2)_{L} \times U(1)_{Y} \times A_{4}$ invariant Yukawa term
$\bar{\ell}_{L}N_{R}\Phi$ is forbidden from the Lagrangian so that the TBM pattern for the neutrino
mixing matrix could be obtained after the contributions from the neutrino and charged lepton sectors are
combined at tree level (i.e., without the dimension-5 operators)~\cite{Ahn:2010cc}.

Taking the $A_{4}$ symmetry breaking scale to be above the electroweak scale in our scenario, i.e.,
$\langle\chi\rangle > \langle \varphi^{0}\rangle$, and assuming the vacuum alignment of the fields
$\langle\chi_{i}\rangle$ as
 \begin{eqnarray}
  \langle \chi_{1} \rangle&\equiv&\upsilon_{\chi}\neq0,
  ~~~~~ \langle \chi_{2} \rangle=\langle\chi_{3}\rangle=0~,
 \label{subgroup}
 \end{eqnarray}
the right-handed Majorana neutrino mass matrix becomes $M$ times the unity matrix plus
a certain matrix having only off-diagonal elements driven by $\langle\chi\rangle$, which can be
expressed as
 \begin{eqnarray}
 M_{R}=M{\left(\begin{array}{ccc}
 1 &  0 &  0 \\
 0 &  1 &  \kappa' e^{i\xi} \\
 0 &  \kappa' e^{i\xi} &  1
 \end{array}\right)}~,
 \label{MR2}
 \end{eqnarray}
where $\kappa'=|\lambda^{s}_{\chi}\upsilon_{\chi}/M|$ with $\langle \chi_{i} \rangle
=\upsilon_{\chi_{i}}~(i=1,2,3)$.
Without loss of generality, we shall assume that the elements of $M_R$ are real: i.e.,
with the definition $\kappa \equiv \kappa' e^{i \xi}$, we obtain $\kappa = \kappa'$ for $\xi = 0$,
and $\kappa = - \kappa'$ for $\xi = \pi$.

After the $A_4$ singlet field $\eta$ acquires the VEV $\langle\eta^{0}\rangle = \upsilon$
which is assumed to be the electroweak scale, the Yukawa interaction terms
~$y_{\nu}\bar{L}_{L}N_{R}\tilde{\eta}$~
and ~$\frac{y^{s, ~a}_{N}}{\Lambda}[(\bar{L}_{L}N_{R})_{{\bf 3}_{s, ~a}}\chi]\cdot\tilde{\eta}$~ can
be combined into the term ~$\upsilon~ \overline{\nu}_{L}Y_{\nu}N_{R}$,~ where the higher energy scale
VEV alignment of the $\chi$ fields given in Eq.~(\ref{subgroup}) has also been used and the neutrino
Yukawa coupling matrix $Y_{\nu}$ is given by
 \begin{eqnarray}
 Y_{\nu}=e^{i\rho}x{\left(\begin{array}{ccc}
 1 &  0 &  0 \\
 0 & 1 & y_{1}e^{i\rho_{1}} \\
 0 &  y_{2}e^{i\rho_{2}} & 1
 \end{array}\right)}~,
 \label{yukawaNu}
 \end{eqnarray}
with $x=|y_{\nu}|$,~ $y_{1}=|y^{s}_{N}+y^{a}_{N}|\upsilon_{\chi}/ (x\Lambda)$,~
$y_{2}=|y^{s}_{N}-y^{a}_{N}|\upsilon_{\chi}/ (x\Lambda)$,~ $\rho = \arg(y_{\nu})$,~
$\rho_1 = \arg(y^{s}_{N} +y^{a}_{N}) -\rho$~ and ~
$\rho_2 = \arg(y^{s}_{N} -y^{a}_{N}) -\rho$.~
Eq.~(\ref{yukawaNu}) indicates that, once the VEV alignment in Eq.~(\ref{subgroup}) is taken, the
$A_{4}$ symmetry is spontaneously broken and its sub-symmetry $Z_{2}$~\cite{He:2006dk} is also
simultaneously broken by the effects of the higher dimensional operators.
Therefore, low energy CP violation responsible for the neutrino oscillation as well as high energy
CP violation responsible for leptogenesis in the neutrino sector can be generated by the off-diagonal
terms of the neutrino Yukawa coupling matrix $Y_{\nu}$~\cite{Ahn:2010cc}.

\subsection{The charged fermion sector}

In the charged fermion sector, the Yukawa interactions $(d\leq5)$ including dimension-5 operators
driven by the $\chi$ field, invariant under $SU(2)_L \times U(1)_Y \times A_{4} \times Z_{2}$,
are given by
 \begin{eqnarray}
 {\cal L}^{f}_{\rm Yuk} = {\cal L}^{u}_{\rm Yuk} +{\cal L}^{d}_{\rm Yuk} +{\cal L}^{\ell}_{\rm Yuk}~,
 \label{lagrangianCh}
 \end{eqnarray}
where
 \begin{eqnarray}
 -{\cal L}^{u}_{\rm Yuk} &=& y_{u}(\bar{Q}_{L}\tilde{\Phi})_{{\bf 1}}u_{R}
  +y_{c}(\bar{Q}_{L}\tilde{\Phi})_{{\bf 1}'}u''_{R}
  +y_{t}(\bar{Q}_{L}\tilde{\Phi})_{{\bf 1}''}u'_{R}\nonumber\\
 &+& \frac{y^{s}_{u}}{\Lambda}[(\bar{Q}_{L}\tilde{\Phi})_{{\bf 3}_{s}}\chi]u_{R}
  +\frac{y^{s}_{c}}{\Lambda}[(\bar{Q}_{L}\tilde{\Phi})_{{\bf 3}_{s}}\chi]_{{\bf 1}'}u''_{R}
  +\frac{y^{s}_{t}}{\Lambda}[(\bar{Q}_{L}\tilde{\Phi})_{{\bf 3}_{s}}\chi]_{{\bf 1}''}u'_{R}\nonumber\\
 &+&\frac{y^{a}_{u}}{\Lambda}[(\bar{Q}_{L}\tilde{\Phi})_{{\bf 3}_{a}}\chi]u_{R}
  +\frac{y^{a}_{c}}{\Lambda}[(\bar{Q}_{L}\tilde{\Phi})_{{\bf 3}_{a}}\chi]_{{\bf 1}'}u''_{R}
  +\frac{y^{a}_{t}}{\Lambda}[(\bar{Q}_{L}\tilde{\Phi})_{{\bf 3}_{a}}\chi]_{{\bf 1}''}u'_{R}
  + {\rm H.c.} ~, \\
 -{\cal L}^{d}_{\rm Yuk} &=&
  y_{d}(\bar{Q}_{L}\Phi)_{{\bf 1}}d_{R}+y_{s}(\bar{Q}_{L}\Phi)_{{\bf 1}'}d''_{R}
  +y_{b}(\bar{Q}_{L}\Phi)_{{\bf 1}''}d'_{R}\nonumber\\
 &+&\frac{y^{s}_{d}}{\Lambda}[(\bar{Q}_{L}\Phi)_{{\bf 3}_{s}}\chi]d_{R}
  +\frac{y^{s}_{s}}{\Lambda}[(\bar{Q}_{L}\Phi)_{{\bf 3}_{s}}\chi]_{{\bf 1}'}d''_{R}
  +\frac{y^{s}_{b}}{\Lambda}[(\bar{Q}_{L}\Phi)_{{\bf 3}_{s}}\chi]_{{\bf 1}''}d'_{R} \nonumber\\
 &+&\frac{y^{a}_{d}}{\Lambda}[(\bar{Q}_{L}\Phi)_{{\bf 3}_{a}}\chi]d_{R}
  +\frac{y^{a}_{s}}{\Lambda}[(\bar{Q}_{L}\Phi)_{{\bf 3}_{a}}\chi]_{{\bf 1}'}d''_{R}
  +\frac{y^{a}_{b}}{\Lambda}[(\bar{Q}_{L}\Phi)_{{\bf 3}_{a}}\chi]_{{\bf 1}''}d'_{R}
  + {\rm H.c.} ~, \\
 -{\cal L}^{\ell}_{\rm Yuk} &=&
  y_{e}(\bar{L}_{L}\Phi)_{{\bf 1}}l_{R}+y_{\mu}(\bar{L}_{L}\Phi)_{{\bf 1}'}l''_{R}
  +y_{\tau}(\bar{L}_{L}\Phi)_{{\bf 1}''}l'_{R}\nonumber\\
 &+&\frac{y^{s}_{e}}{\Lambda}[(\bar{L}_{L}\Phi)_{{\bf 3}_{s}}\chi]l_{R}
  +\frac{y^{s}_{\mu}}{\Lambda}[(\bar{L}_{L}\Phi)_{{\bf 3}_{s}}\chi]_{{\bf 1}'}l''_{R}
  +\frac{y^{s}_{\tau}}{\Lambda}[(\bar{L}_{L}\Phi)_{{\bf 3}_{s}}\chi]_{{\bf 1}''}l'_{R}\nonumber\\
 &+&\frac{y^{a}_{e}}{\Lambda}[(\bar{L}_{L}\Phi)_{{\bf 3}_{a}}\chi]l_{R}
  +\frac{y^{a}_{\mu}}{\Lambda}[(\bar{L}_{L}\Phi)_{{\bf 3}_{a}}\chi]_{{\bf 1}'}l''_{R}
  +\frac{y^{a}_{\tau}}{\Lambda}[(\bar{L}_{L}\Phi)_{{\bf 3}_{a}}\chi]_{{\bf 1}''}l'_{R}
  + {\rm H.c.} ~,
 \end{eqnarray}
with $\tilde{\Phi}\equiv i\tau_{2}\Phi^{\ast}$.
In the above Lagrangian, each charged fermion sector has three independent Yukawa terms, all involving
the $A_{4}$-triplet Higgs field $\Phi$. The left-handed quark and lepton doublets $Q_{L}$ and $L_{L}$
transform as a triplet ${\bf 3}$, while the right-handed quarks and leptons $(u_{R}, d_{R}, e_{R}),
(c_{R}, s_{R}, \mu_{R}), (t_{R}, b_{R}, \tau_{R})$ transform as ${\bf 1}$, ${\bf 1}''$ and ${\bf 1}'$,
respectively.
We note that the $A_4$-triplet scalar field $\chi$ drives the dimension-5 operators
both in the neutrino sector shown in Eq.~(\ref{lagrangian}) and in the charged fermion sector
shown in Eq.~(\ref{lagrangianCh}).
Thus, this $\chi$ field plays a role to connect the neutrino, charged lepton and quark sectors
to one another through the higher dimensional operators.

We assume that the VEVs of the $A_{4}$-triplet $\Phi$ can be equally aligned, that is,
$\langle \varphi^{0} \rangle = (\upsilon,\upsilon,\upsilon)$, with the VEV alignment in
Eq.~(\ref{subgroup}). Then the charged fermion mass matrix $m_f$ can be explicitly expressed as
 \begin{eqnarray}
 m_{f}&=&U_{\omega}\sqrt{3}{\left(\begin{array}{ccc}
 m^{f}_{11} & m^{f}_{12} & m^{f}_{13} \\
 m^{f}_{21} & m^{f}_{22} & m^{f}_{23} \\
 m^{f}_{31} & m^{f}_{32} & m^{f}_{33}
 \end{array}\right)}~,~~~~~~~~~~~~~ {\rm with}~~
 U_{\omega}=\frac{1}{\sqrt{3}}{\left(\begin{array}{ccc}
 1 &  1 &  1 \\
 1 &  \omega &  \omega^{2} \\
 1 &  \omega^{2} &  \omega
 \end{array}\right)}\nonumber\\
 &=& U_{\omega}V^{f}_{L} ~{\rm Diag}(m_{f_{1}},m_{f_{2}},m_{f_{3}})~ V^{f\dag}_{R}~,
 \label{CHcorrect}
 \end{eqnarray}
where $f$ denotes the charged lepton, up- or down-type quarks.
$U_{\omega}V^{f}_{L}$ and $V^{f}_{R}$ are the diagonalization matrices for $m_f$.
The elements of $m_f$ are given by
 \begin{eqnarray}
  m^{f}_{11}&=&\upsilon(y_{f_{1}}+2h_{1}/3) ,~~m^{f}_{12}=2\upsilon h_{2}/3,~~~~~~~~~~~~~~~~~~~
  m^{f}_{13}=2\upsilon h_{3}/3~, \nonumber\\ m^{f}_{21}&=&\upsilon(g_{1}-h_{1})/3 ,~~~~~
  m^{f}_{22}=\upsilon(y_{f_{2}}+(g_{2}-h_{2})/3),~~m^{f}_{23}=\upsilon(g_{3}-h_{3})/3~,\nonumber\\
  m^{f}_{31}&=&-\upsilon(g_{1}+h_{1})/3 ,~~~m^{f}_{32}=-\upsilon(g_{2}+h_{2})/3,~~~~~~~~~
  m^{f}_{33}=\upsilon(y_{f_{3}}-(g_{3}+h_{3})/3)~,
 \label{mf_elements}
 \end{eqnarray}
where the complex parameters $h_i$ and $g_i$ are defined as
~$h_{1}=\upsilon_{\chi}y^{s}_{f_{1}}/\Lambda$, ~$h_{2}=\upsilon_{\chi}y^{s}_{f_{2}}/\Lambda$,
~$h_{3}=\upsilon_{\chi}y^{s}_{f_{3}}/\Lambda$,
~$g_{1}=-i\sqrt{3}\upsilon_{\chi}y^{a}_{f_{1}}/\Lambda$,
~$g_{2}=-i\sqrt{3}\upsilon_{\chi}y^{a}_{f_{2}}/\Lambda$,
~$g_{3}=-i\sqrt{3}\upsilon_{\chi}y^{a}_{f_{3}}/\Lambda$.
By taking the VEV alignment of $\langle \chi_{i} \rangle$ given in Eq.~(\ref{subgroup}) with the equal
VEV alignment of $\langle \phi^{0} \rangle$, the $A_{4}$ symmetry is  spontaneously broken and at the
same time its sub-symmetry $C_{3}$ is also broken through the dimension-5 operators~\cite{He:2006dk}.
One of the most interesting features observed by experiments on the charged fermions is that the mass
spectra of quarks and charged leptons are strongly hierarchical, {\it i.e.}, the masses of the third
generation fermions are much heavier than those of the first and second generation fermions.
For the elements of $m_f$ given in Eq.~(\ref{mf_elements}), taking into account the most natural case
that the charged fermion Yukawa couplings have the strong hierarchy ~$y_{f_{3}} \gg y_{f_{2}}
\gg y_{f_{1}}$~ and the off-diagonal elements generated by the higher dimensional operators are
generally smaller in magnitude than the diagonal ones, we make a plausible assumption
 \begin{eqnarray}
  y_{f_{3}} &\gg& |g_{3}| \sim ~({\rm or}\gg)~ |h_{3}| ~,~~~
  y_{f_{2}} \gg |g_{2}| \sim |h_{2}|~,~~~ y_{f_{1}} \gg |g_{1}| \sim |h_{1}| ~, \nonumber\\
  |h_{3}| &\sim& ({\rm or}\gg)~|h_{2}|~,~~~|h_{2}|\sim({\rm or}\gg)~y_{f_{1}} ~.
 \label{hierarchy}
 \end{eqnarray}
Then $V^{f}_{L}$ and $V^{f}_{R}$ can be determined by diagonalizing the matrices
$U^{\dag}_{\omega}m_{f}m^{\dag}_{f}U_{\omega}$ and $m^{\dag}_{f}m_{f}$, respectively, indicated from
Eq.~(\ref{CHcorrect}). Especially, the mixing matrix $V^{f}_{L}$ becomes one of the matrices composing
the Pontecorvo-Maki-Nakagawa-Sakata (PMNS) and CKM ones and it can be approximated, due to the
strong hierarchy expressed in Eq.~(\ref{hierarchy}), as
 \begin{eqnarray}
 \footnotesize
 V^{f}_{L}\simeq
 {\left(\begin{array}{ccc}
 1 - \frac{1}{2} \left| \frac{m^{f}_{12}}{m^{f}_{22}} \right|^{2}
  & \left| \frac{m^{f}_{12}}{m^{f}_{22}} \right| e^{i\phi^{f}_{3}}
  & \left| \frac{m^{f}_{13}}{m^{f}_{33}} \right| e^{i\phi^{f}_{2}}  \\
 - \left| \frac{m^{f}_{12}}{m^{f}_{22}} \right| e^{-i\phi^{f}_{3}}
  & 1 -\frac{1}{2} \left| \frac{m^{f}_{12}}{m^{f}_{22}} \right|^{2}
  & \left| \frac{m^{f}_{23}}{m^{f}_{33}} \right| e^{i\phi^{f}_{1}} \\
 - \left| \frac{m^{f}_{13}}{m^{f}_{33}} \right| e^{-i\phi^{f}_{2}}
   + \left| \frac{m^{f}_{12}}{m^{f}_{22}} \right|
   \left| \frac{m^{f}_{23}}{m^{f}_{33}} \right| e^{-i(\phi^{f}_{3}+\phi^{f}_{1})}
  & - \left| \frac{m^{f}_{23}}{m^{f}_{33}} \right| e^{-i\phi^{f}_{1}}
   - \left| \frac{m^{f}_{13}}{m^{f}_{33}} \right|
    \left| \frac{m^{f}_{12}}{m^{f}_{22}} \right| e^{i(\phi^{f}_{3}-\phi^{f}_{2})}
  & 1
 \end{array}\right)}
 \label{mixingL}
 \end{eqnarray}
where we have used  ~$|m^{f}_{12} /m^{f}_{22}|, |m^{f}_{13} /m^{f}_{33}|, |m^{f}_{23} /m^{f}_{33}| \ll 1$,~
and ~$\phi^{f}_{1} \simeq \arg \Big( m^{f}_{22}m^{f\ast}_{32} +m^{f}_{23}m^{f\ast}_{33} \Big) /2$,~
$\phi^{f}_{2} \simeq \arg \Big( m^{f}_{11}m^{f\ast}_{31} +m^{f}_{13}m^{f\ast}_{33} \Big) /2$~
and ~$\phi^{f}_{3} \simeq \arg \Big( m^{f}_{11}m^{f\ast}_{21} +m^{f}_{12}m^{f\ast}_{22} \Big) /2$.

There exist several empirical fermion mass ratios in the charged lepton, up- and down-type quark sectors
calculated from the measured values~\cite{pdg} :
 \begin{eqnarray}
  \frac{m_{e}}{m_{\tau}} &\simeq&  2.9\times10^{-4}~,~~~~~~~~
  \frac{m_{\mu}}{m_{\tau}} \simeq 5.9\times10^{-2}~, \nonumber\\
  \frac{m_{d}}{m_{b}} &\simeq&  1.2\times 10^{-3}~,~~~~~~~~
  \frac{m_{s}}{m_{b}} \simeq 2.4\times10^{-2}~, \nonumber\\
  \frac{m_{u}}{m_{t}} &\simeq& 1.4\times 10^{-5}~,~~~~~~~~
  \frac{m_{c}}{m_{t}}\simeq7.4\times10^{-3}~,
 \label{massRatio}
 \end{eqnarray}
which imply that the possible quark-lepton symmetry~\cite{Pati:1973rp} is broken by the masses of quarks
and leptons. Thus, it is not expected that the known quark mixing pattern is transmitted to the lepton
sector in the exactly same form. Nevertheless, a key point inferred from Eq.~(\ref{massRatio}) is
that the mass spectrum of the charged leptons exhibits a similar hierarchical pattern to that of the
down-type quarks, unlike that of the up-type quarks which shows a much stronger hierarchical pattern.
For instance,  in terms of the Cabbibo angle $\lambda \equiv \sin\theta_{\rm C} \approx |V_{us}|$, the
fermion masses scale as ~$(m_{e},m_{\mu}) \approx (\lambda^{5},\lambda^{2})~ m_{\tau}$,~
$(m_{d},m_{s}) \approx (\lambda^{4},\lambda^{2})~ m_{b}$~ and
~$(m_{u},m_{c}) \approx (\lambda^{8},\lambda^{4})~ m_{t}$,~ which may represent the following two
facts: (i) the CKM matrix is mainly generated by the mixing matrix of the down-type quark sector, and
(ii) the mixing matrix of the charged lepton sector is similar to that of the down-type quark sector,
when the Lagrangian~(\ref{lagrangianCh}) is also taken into account.
Further, there is another interesting empirical relation
 \begin{eqnarray}
  |V_{us}| \approx \left( \frac{m_{d}}{m_{s}} \right)^{\frac{1}{2}}
  \approx 3 \left(\frac{m_{e}}{m_{\mu}} \right)^{\frac{1}{2}}~,
 \label{Vus}
 \end{eqnarray}
which has been known for quite a long time~\cite{Gatto:1968ss}.

\subsubsection{The up-type quark sector and its mixing matrix}

From Eq.~(\ref{CHcorrect}) we see that the up-type quark mass matrix $m_{f_{u}}$ can be diagonalized
in the mass basis by a biunitary transformation,
$V^{u\dag}_{L}U^{\dag}_{\omega}m_{f_{u}}V^{u}_{R}={\rm Diag}(m_{u},m_{c},m_{t})$.
The matrices $V^{u}_{L}$ and $V^{u}_{R}$ can be determined by diagonalizing the matrices
$U^{\dag}_{\omega}m_{f_{u}}m^{\dag}_{f_{u}}U_{\omega}$ and $m^{\dag}_{f_{u}}m_{f_{u}}$, respectively.
Especially, the left-handed up-type quark mixing matrix $V^{u}_{L}$ becomes one of the matrices
composing the CKM matrix such as $V_{\rm CKM} \equiv V^{d\dag}_{L}V^{u}_{L}$ (see Eq.~(\ref{ckm1})
below).
Due to the measured value of $m_u / m_t$ in Eq.~(\ref{massRatio}), it is impossible to generate the
Cabbibo angle, $\lambda\approx|V_{us}|$, from the mixing between the first and second generations in
the up-type quark sector: if one sets ~$|(V^{u}_{L})_{12}| = |m^{u}_{12}/m^{u}_{22}|\approx\lambda$,~
then from Eq.~(\ref{hierarchy}) one obtains
~$m_{u}/m_{t}\approx|m^{u}_{12}/m^{u}_{22}|~ |m^{u}_{22}/m^{u}_{33}|\approx\lambda^{5}$,~ in
discrepancy with the measured $m_{u}/m_{t}\approx\lambda^{8}$ in Eq.~(\ref{massRatio}).
To determine the correct up-type quark mixing matrix, using both Eqs.~(\ref{hierarchy})
and (\ref{massRatio}), we obtain $m_{c}/m_{t}\approx|m^{u}_{22}/m^{u}_{33}|\approx \lambda^{4}$,
$m_{u}/m_{c}\approx|m^{u}_{11}/m^{u}_{22}|\approx \lambda^{4}$ and
$m_{u}/m_{t}\approx|m^{u}_{11}/m^{u}_{33}|\approx \lambda^{8}$.
Then the up-type quark mixing matrix $V^u_L$ can be approximated as
 \begin{eqnarray}
 V^{u}_{L}\simeq
{\left(\begin{array}{ccc}
 1 & \lambda^{4}e^{i\phi^{u}_{3}}  & \lambda^{4}e^{i\phi^{u}_{2}}  \\
 - \lambda^{4}e^{-i\phi^{u}_{3}} &  1 & \lambda^{4}e^{i\phi^{u}_{1}} \\
 - \lambda^{4}e^{-i\phi^{u}_{2}} &  - \lambda^{4}e^{-i\phi^{u}_{1}} & 1
 \end{array}\right)} +{\cal O}(\lambda^{5}) ~,
 \label{UL}
 \end{eqnarray}
which indicates that the mixing in the up-type quark sector does not affect the leading order
contributions in $\lambda$.  It leads to the fact that the Cabbibo angle should arise from the mixing
between the first and second generations in the down-type quark sector.

\subsubsection{The down-type quark sector and its mixing matrix}

The empirical relation (\ref{Vus}) shows that the mass hierarchy of the down-type quark
sector is similar to that of the charged-lepton one.  Now let us consider the down-type quark sector
to obtain the realistic CKM matrix. From Eq.~(\ref{hierarchy}) and the measured down-type quark mass
hierarchy in Eq.~(\ref{massRatio}), we find
~$m_{s}/m_{b}\approx|m^{d}_{22}/m^{d}_{33}| \approx 0.6 \,\lambda^{2}$,
~$m_{d}/m_{b}\approx|m^{d}_{11}/m^{d}_{33}| \approx 0.7 \,\lambda^{4}$~ and
~$m_{d}/m_{s}\approx|m^{d}_{11}/m^{d}_{22}| \approx \lambda^{2}$.~
From Eqs.~(\ref{hierarchy}) and (\ref{mixingL}), we obtain
$|(V^{d}_{L})_{12}| \approx |m^{d}_{12}/m^{d}_{22}| \approx 1.7 \,\lambda^{-2} |m^{d}_{12}/m^{d}_{33}|$,
which means ~$|m^{d}_{12}/m^{d}_{33}| \approx 0.6 \,\lambda^{3}$~ for $|(V^{d}_{L})_{12}|
\approx \lambda$.~  In order to get the correct CKM matrix element
$|m^{d}_{13} / m^{d}_{33}| \sim {\cal O}(\lambda^{3})$, we need to make an additional assumption:
from Eq.~(\ref{hierarchy}) the hierarchy normalized by the bottom quark mass can be
expressed as
 \begin{eqnarray}
  1\gg \frac{|m^{d}_{22}|}{|m^{d}_{33}|}\sim \frac{|m^{d}_{23}|}{|m^{d}_{33}|}
  \gg \frac{|m^{d}_{13}|}{|m^{d}_{33}|}\sim \frac{|m^{d}_{12}|}{|m^{d}_{33}|}
  \gg \frac{|m^{d}_{11}|}{|m^{d}_{33}|}\sim \frac{|m^{d}_{32}|}{|m^{d}_{33}|}
  \gg \frac{|m^{d}_{21}|}{|m^{d}_{33}|}\sim \frac{|m^{d}_{31}|}{|m^{d}_{33}|} ~.
 \label{hierarchyD}
 \end{eqnarray}
Then, we can obtain the mixing matrix $V^d_L$ of the down-type quarks: under the constraint of unitarity,
it can be written as
 \begin{eqnarray}
 V^{d}_{L}\simeq
 {\left(\begin{array}{ccc}
 1 -\frac{\lambda^{2}}{2} & \lambda e^{i\phi^{d}_{3}}  & A' \lambda^{3} e^{i\phi^{d}_{2}}  \\
 - \lambda e^{-i\phi^{d}_{3}} &  1-\frac{\lambda^{2}}{2} & A \lambda^{2} e^{i\phi^{d}_{1}} \\
 -A' \lambda^{3} e^{-i\phi^{d}_{2}} +A \lambda^{3} e^{-i(\phi^{d}_{3}+\phi^{d}_{1})}
  &  -A \lambda^{2} e^{-i\phi^{d}_{1}} & 1
 \end{array}\right)}+{\cal O}(\lambda^{4})~,
 \label{DL}
 \end{eqnarray}
where $A$ and $A'$ are positive real numbers of order unity.  Later in Eq.~(\ref{ckm1}), we shall see
that this form of $V^d_L$ indeed becomes the realistic CKM matrix.

\subsubsection{The charged lepton sector and its mixing matrix}

Now let us turn to the charged lepton sector.  From Eq.~(\ref{hierarchy}) and the measured charged
lepton mass hierarchy in Eq.~(\ref{massRatio}), we obtain
~$m_{\mu}/m_{\tau} \approx |m^{\ell}_{22}/m^{\ell}_{33}| \approx \lambda^{2}$,
~$m_{e}/m_{\tau} \approx |m^{\ell}_{11}/m^{\ell}_{33}| \approx 0.6 \,\lambda^{5}$~ and
~$m_{e}/m_{\mu} \approx |m^{\ell}_{11}/m^{\ell}_{22}| \approx 0.5 \,\lambda^{3}$.~
Similarly to the case of the down-type quark sector, from Eqs.~(\ref{hierarchy}) and
(\ref{mixingL}), we find $|(V^{\ell}_{L})_{12}| \approx |m^{\ell}_{12}/m^{\ell}_{22}| \approx
\lambda^{-2} |m^{\ell}_{12}/m^{\ell}_{33}|$, which leads to ~$|m^{\ell}_{12}/m^{\ell}_{33}| \approx
\lambda^3$~ for $|(V^{\ell}_{L})_{12}| \approx \lambda$.
With the hierarchy among the couplings given in Eq.~(\ref{hierarchy}) and under the constraint of
unitarity, we obtain three different types of the mixing matrix $V^{\ell}_L$ for the charged
leptons, which we will call {\it Scenario-I, -II, -III,} respectively, as below.

First, if we set the condition $|m^{\ell}_{23}|\gg|m^{\ell}_{13}|$ similarly to Eq.~(\ref{DL}), from
the hierarchy shown in Eq.~(\ref{hierarchy}), the mass hierarchy normalized by the tau mass
$m_{\tau}\approx|m^{\ell}_{33}|$ is obtained as
 \begin{eqnarray}
  1 \gg \frac{|m^{\ell}_{22}|}{|m^{\ell}_{33}|}\sim \frac{|m^{\ell}_{23}|}{|m^{\ell}_{33}|}
  \gg \frac{|m^{\ell}_{13}|}{|m^{\ell}_{33}|}\sim \frac{|m^{\ell}_{12}|}{|m^{\ell}_{33}|}
  \gg \frac{|m^{\ell}_{11}|}{|m^{\ell}_{33}|}\sim \frac{|m^{\ell}_{32}|}{|m^{\ell}_{33}|}
  \gg \frac{|m^{\ell}_{21}|}{|m^{\ell}_{33}|}\sim \frac{|m^{\ell}_{31}|}{|m^{\ell}_{33}|} ~.
 \label{hierarchyL}
 \end{eqnarray}
The resulting mixing matrix $V^{\ell}_L$ ({\it Scenario-I}) is given by
 \begin{eqnarray}
 V^{\ell}_{L}\simeq {\left(\begin{array}{ccc}
 1-\frac{\lambda^{2}}{2} & \lambda e^{i\phi^{\ell}_{3}}  & A_{1}\lambda^{3}e^{i\phi^{\ell}_{2}}  \\
 -\lambda e^{-i\phi^{\ell}_{3}} &  1-\frac{\lambda^{2}}{2} & B_{1}\lambda^{2}e^{i\phi^{\ell}_{1}} \\
 -A_{1}\lambda^{3}e^{-i\phi^{\ell}_{2}}+B_{1}\lambda^{3}e^{-i(\phi^{\ell}_{1}+\phi^{\ell}_{3})}
 & -B_{1}\lambda^{2}e^{-i\phi^{\ell}_{1}} & 1
 \end{array}\right)}+{\cal O}(\lambda^{4})~,
 \label{LL}
 \end{eqnarray}
where $A_{1}$ and $B_{1}$ are real and positive ${\cal O}(1)$ coefficients. It is quite similar in form
to the mixing matrix $V^d_L$ of the down-type quarks given in Eq.~(\ref{DL}).

Secondly, if we assign the condition $|m^{\ell}_{12}|\gg|m^{\ell}_{11}|$ in addition to
Eq.~(\ref{hierarchy}), we obtain the mass hierarchy relation normalized by the tau mass:
 \begin{eqnarray}
  1\gg \frac{|m^{\ell}_{22}|}{|m^{\ell}_{33}|}\sim \frac{|m^{\ell}_{23}|}{|m^{\ell}_{33}|}
  \sim\frac{|m^{\ell}_{13}|}{|m^{\ell}_{33}|}\gg \frac{|m^{\ell}_{12}|}{|m^{\ell}_{33}|}
  \gg \frac{|m^{\ell}_{11}|}{|m^{\ell}_{33}|}\sim \frac{|m^{\ell}_{32}|}{|m^{\ell}_{33}|}
  \gg \frac{|m^{\ell}_{21}|}{|m^{\ell}_{33}|}\sim \frac{|m^{\ell}_{31}|}{|m^{\ell}_{33}|}~.
 \label{hierarchyL1}
 \end{eqnarray}
Subsequently, under the unitarity constraint, the mixing matrix $V^{\ell}_L$ ({\it Scenario-II}) can
be recast to
 \begin{eqnarray}
 \small
 V^{\ell}_{L}\simeq {\left(\begin{array}{ccc}
 1 -\frac{\lambda^{2}}{2} & \lambda e^{i\phi^{\ell}_{3}}  & A_{2}\lambda^{2}e^{i\phi^{\ell}_{2}}  \\
 -\lambda e^{-i\phi^{\ell}_{3}} &  1-\frac{\lambda^{2}}{2} & B_{2}\lambda^{2}e^{i\phi^{\ell}_{1}} \\
 -A_{2}\lambda^{2}e^{-i\phi^{\ell}_{2}}+B_{2}\lambda^{3}e^{-i(\phi^{\ell}_{3}+\phi^{\ell}_{1})}
  & -B_{2}\lambda^{2}e^{-i\phi^{\ell}_{1}}-A_{2}\lambda^{3}e^{i(\phi^{\ell}_{3}-\phi^{\ell}_{2})}
  & 1
 \end{array}\right)}+{\cal O}(\lambda^{4})~,
 \label{LL1}
 \end{eqnarray}
where $A_{2}$ and $B_{2}$ are real and positive ${\cal O}(1)$ coefficients.

Finally, adding the assumption of $|m^{\ell}_{13}|\gg|m^{\ell}_{23}|$ to Eq.~(\ref{hierarchy}), we get
the mass hierarchy relation
 \begin{eqnarray}
  1\gg \frac{|m^{\ell}_{22}|}{|m^{\ell}_{33}|}\sim \frac{|m^{\ell}_{13}|}{|m^{\ell}_{33}|}
  \gg\frac{|m^{\ell}_{23}|}{|m^{\ell}_{33}|}\sim \frac{|m^{\ell}_{12}|}{|m^{\ell}_{33}|}
  \gg \frac{|m^{\ell}_{11}|}{|m^{\ell}_{33}|}\sim \frac{|m^{\ell}_{32}|}{|m^{\ell}_{33}|}
  \gg \frac{|m^{\ell}_{21}|}{|m^{\ell}_{33}|}\sim \frac{|m^{\ell}_{31}|}{|m^{\ell}_{33}|}~,
 \label{hierarchyL2}
 \end{eqnarray}
which under the unitarity condition, leads to the mixing matrix $V^{\ell}_L$ ({\it Scenario-III})
 \begin{eqnarray}
 V^{\ell}_{L}\simeq {\left(\begin{array}{ccc}
 1-\frac{\lambda^{2}}{2} & \lambda e^{i\phi^{\ell}_{3}}  & A_{3}\lambda^{2}e^{i\phi^{\ell}_{2}}  \\
 -\lambda e^{-i\phi^{\ell}_{3}} &  1-\frac{\lambda^{2}}{2} & 0 \\
 -A_{3}\lambda^{2}e^{-i\phi^{\ell}_{2}} &  -A_{3}\lambda^{3}e^{i(\phi^{\ell}_{3}-\phi^{\ell}_{2})} & 1
 \end{array}\right)}+{\cal O}(\lambda^{4})~,
 \label{LL2}
 \end{eqnarray}
where $A_{3}$ is a real and positive ${\cal O}(1)$ number.

\subsection{Masses and mixings of quarks and charged leptons}

In the weak eigenstate basis, the Yukawa interactions in Eq.~(\ref{lagrangian}) and the charged gauge
interactions can be written as
 \begin{eqnarray}
 -{\cal L} &=& \frac{1}{2} (\overline{N_{R}})^c M_{R}N_{R}+\overline{\ell_{L}}m_{\ell}\ell_{R}
  + \overline{\nu_{L}}m_{D}N_{R}
  +\frac{g}{\sqrt{2}}W^{-}_{\mu} ~\overline{\ell_{L}}\gamma^{\mu}\nu_{L} \nonumber\\
 &+& \overline{q^{u}_{L}}m_{u}q^{u}_{R}+\overline{q^{d}_{L}}m_{d}q^{d}_{R}
  +\frac{g}{\sqrt{2}}W^{-}_{\mu} ~\overline{q^{d}_{L}}\gamma^{\mu}q^{u}_{L} + {\rm H.c.} ~,
 \label{lagrangianA}
 \end{eqnarray}
with the Dirac neutrino mass $m_{D}=\upsilon Y_{\nu}$.~
From Eq.~(\ref{lagrangianA}) the neutrino mass terms are given by
 \begin{eqnarray}
  -{\cal L}_{\nu}= \frac{1}{2}\overline{n_{L}} ~{\cal M}_{\nu} ~(n_{L})^c + {\rm H.c.} ~,
 \end{eqnarray}
where
 \begin{eqnarray}
  n_{L}={\left(\begin{array}{c}
   \nu_{L} \\
   (N_{R})^c
 \end{array}\right)}~,~~~
  {\cal M}_{\nu}={\left(\begin{array}{cc}
   0 & m_{D} \\
   m^{T}_{D} & M_{R}
 \end{array}\right)} ~.
 \label{numass}
 \end{eqnarray}
Since $M_{R} \gg m_{D}$, the light neutrino mass matrix $m_{\nu}$ at low energies reads
 \begin{eqnarray}
  m_{\nu}&=&-m_{D}M^{-1}_{R}m^{T}_{D} = U_{\nu} ~{\rm Diag}(m_{1},m_{2},m_{3})~ U^{T}_{\nu}
   = -e^{2i\rho}m_{0}
   {\left(\begin{array}{ccc}
   1 & 0 & 0 \\
   0 & A & G  \\
   0 & G & B
   \end{array}\right)} ~,
 \end{eqnarray}
where $m_i ~(i = 1,2,3)$ are the light neutrino mass eigenvalues and
 \begin{eqnarray}
  && m_{0} = \frac{x^{2}\upsilon^{2}}{M} ~,  \hspace{3.5cm}
   A = \frac{1-2y_{1}\kappa e^{i\rho_{1}}+y^{2}_{1}e^{2i\rho_{1}}}{1-\kappa^{2}}~, \nonumber \\
  && B = \frac{1-2y_{2}\kappa e^{i\rho_{2}}+y^{2}_{2}e^{2i\rho_{2}}}{1-\kappa^{2}}~,~~~~~
   G = \frac{\kappa-y_{1}e^{i\rho_{1}}-y_{2}e^{i\rho_{2}} +\kappa y_{1}y_{2}e^{i\tilde{\rho}_{12}}}
    {\kappa^{2}-1}~,
 \end{eqnarray}
with $\tilde{\rho}_{12}\equiv\rho_{1}+\rho_{2}$.~
The parameters $y_{1,2}$ and the phases $\rho$ and $\rho_{1,2}$ have been defined in
Eq.~(\ref{yukawaNu}).  The diagonalization matrix $U_{\nu}$ of the light neutrino mass matrix $m_{\nu}$
is given by
 \begin{eqnarray}
 U_{\nu}=e^{i\pi/2}{\left(\begin{array}{ccc}
 1 & 0 & 0 \\
 0 & e^{i\varphi_{1}}  &  0 \\
 0 & 0  &  e^{i\varphi_{2}}
 \end{array}\right)}{\left(\begin{array}{ccc}
 0 & 1 & 0 \\
 \cos\theta & 0  &  -\sin\theta \\
 \sin\theta & 0  &  \cos\theta
 \end{array}\right)}{\left(\begin{array}{ccc}
 e^{i\xi_{1}} & 0 & 0 \\
 0 & e^{i\xi_{2}}  &  0 \\
 0 & 0  &  e^{i\xi_{3}}
 \end{array}\right)} ~,
 \label{mixing}
 \end{eqnarray}
where the Majorana phases $\xi_{i}$ can be absorbed into the neutrino mass eigenstate fields.  The
phases $\varphi_{21}$ and the mixing angle $\theta$ are given by
 \begin{eqnarray}
  \varphi_{21}\equiv\varphi_{2}-\varphi_{1}=\arg(GA^{\ast}+BG^{\ast}) ~,~~~~
  \tan2\theta=\frac{2|AG^{\ast}+GB^{\ast}|}{|A|^{2}-|B|^{2}}~,
  \label{Phi21Theta}
 \end{eqnarray}
which indicate that, in the limit of $y_{1,2}$ approaching to zero, the angle $\theta$ goes to $\pm\pi/4$
and the phase $\varphi_{21}$ goes to $0(\pi)$ for the negative (positive) sign of $\kappa$, due to
 \begin{eqnarray}
  \cos\varphi_{21} \approx \frac{-\kappa+(1+\kappa^{2})(y_{1}\cos\rho_{1}+y_{2}\cos\rho_{2})}
   {\sqrt{\kappa^{2}-2\kappa(1+\kappa^{2})(y_{1}\cos\rho_{1}+y_{2}\cos\rho_{2})}}~.
  \label{Cphi21}
 \end{eqnarray}
Interestingly enough, as we can see in the second relation of Eq.~(\ref{Phi21Theta}), the sign of
$\theta$ depends on the relative size between $|A|^{2}$ and $|B|^{2}$, which is in a good approximation
given by
 \begin{eqnarray}
  |A|^{2}-|B|^{2}\simeq\frac{4\kappa(y_{2}\cos\rho_{2}-y_{1}\cos\rho_{1})}{(1-\kappa^{2})^{2}}~.
  \label{AB}
 \end{eqnarray}
The above relation is important to determine the patterns of the mass spectrum, as we shall see in
Eq.~(\ref{deltam2}). As will be shown later, for $y_{1,2}\ll1$, the values of $\theta=\pi/4+\delta$
and $\theta=-\pi/4+\delta$ with $\delta\ll1$ correspond to $\cos\varphi_{21}>0$ and
$\cos\varphi_{21}<0$, respectively, which are constrained by the experimental data for the neutrino
mixing angles. The light neutrino mass eigenvalues are obtained as
 \begin{eqnarray}
  m^{2}_{1}&=& m^{2}_{0} ~\Big( |A|^{2}\cos^{2}\theta+|B|^{2}\sin^{2}\theta+|G|^{2}
   +|AG^{\ast} +GB^{\ast}|\sin2\theta \Big) ~, \nonumber\\
  m^{2}_{2}&=& m^{2}_{0} ~, \nonumber\\
  m^{2}_{3}&=& m^{2}_{0} ~\Big( |A|^{2}\sin^{2}\theta+|B|^{2}\cos^{2}\theta+|G|^{2}
   -|AG^{\ast}+GB^{\ast}|\sin2\theta \Big) ~.
 \label{mass2}
 \end{eqnarray}
Because of the observed hierarchy $|\Delta m^{2}_{\rm atm}|\gg\Delta m^{2}_{\rm sol}>0$, and the
requirement of Mikheyev-Smirnov-Wolfenstein resonance for solar neutrinos, there are two possible
neutrino mass spectrum: (i) $m_{1}<m_{2}<m_{3}$ (normal mass spectrum) which corresponds to
$\theta=-\pi/4+\delta$ and (ii) $m_{3}<m_{1}<m_{2}$ (inverted mass spectrum) corresponding to
$\theta=\pi/4+\delta$. The solar and atmospheric mass-squared differences are given by
\begin{eqnarray}
 \Delta m^{2}_{\rm sol}\equiv m^{2}_{2}-m^{2}_{1}
  &=& m^{2}_{0} \Big( 1-|G|^{2}-|A|^{2}\cos^{2}\theta-|B|^{2}\sin^{2}\theta
   +|AG^{\ast}+GB^{\ast}|\sin2\theta \Big) ~,\nonumber\\
 \Delta m^{2}_{\rm atm}\equiv m^{2}_{3}-m^{2}_{1}
  &=& -2m^{2}_{0}\frac{|AG^{\ast}+GB^{\ast}|}{\sin2\theta}~,
 \label{deltam2}
\end{eqnarray}
which are constrained by the neutrino oscillation experimental results.

On the other hand, from Eq.~(\ref{lagrangianA}), to diagonalize the charged fermion mass matrices such
that
 \begin{eqnarray}
  V^{f\dag}_{L} U^{\dag}_{\omega} ~m_{f}~ V^{f}_{R} = {\rm Diag}(m_{f_{1}},m_{f_{2}},m_{f_{3}})~
  \equiv \hat m_f  ~,
 \end{eqnarray}
we can rotate the fermion fields from the weak eigenstates to the mass eigenstates:
 \begin{eqnarray}
  && \ell_{L}\rightarrow V^{\ell ~\dag}_{L}U^{\dag}_{\omega}\ell_{L}~,~~~~~~
   \ell_{R}\rightarrow V^{\dag}_{R}\ell_{R}~,~~~~~~
   \nu_{L}\rightarrow U^{\dag}_{\nu}\nu_{L}~,\nonumber\\
  && q^{u(d)}_{L} \rightarrow V^{u(d) \dag}_{L} U^{\dag}_{\omega}~ q^{u(d)}_{L} ~,~~~~~~
   q^{u(d)}_{R} \rightarrow V^{u(d) ~T}_{R} q^{u(d)}_{R} ~.
\label{basis}
 \end{eqnarray}
Then, from the charged current terms in Eq.~(\ref{lagrangianA}), we obtain the CKM and PMNS matrices
 \begin{eqnarray}
  V_{\rm CKM} &=& \Big( U_{\omega}V^{d}_{L} \Big)^{\dag} \Big( U_{\omega}V^{u}_{L} \Big)
   = V^{d\dag}_{L}V^{u}_{L}~,~~~~~~~~
  U_{\rm PMNS} = V^{\ell ~\dag}_{L}U^{\dag}_{\omega}U_{\nu}~.
 \label{mixing relation}
 \end{eqnarray}
From Eqs.~(\ref{UL}) and (\ref{DL}), if we set
 \begin{eqnarray}
  A' e^{i\phi^{d}_{2}} = A (\rho +i\eta)~, ~~~~~ \phi^{d}_{1} = \phi^{d}_{3}= 0 ~,
 \end{eqnarray}
then we obtain the CKM matrix in the Wolfenstein parametrization~\cite{Wolfenstein:1983yz} given by
 \begin{eqnarray}
  V_{\rm CKM}=V^{d\dag}_{L}V^{u}_{L}\simeq V^{d\dag}_{L}
  \simeq{\left(\begin{array}{ccc}
 1-\frac{\lambda^{2}}{2} & \lambda  & A\lambda^{3}(\rho+i\eta)  \\
 -\lambda &  1-\frac{\lambda^{2}}{2} & A\lambda^{2} \\
 A\lambda^{3}(1-\rho+i\eta) & -A\lambda^{2}  & 1
 \end{array}\right)}+{\cal O}(\lambda^{4})~.
 \label{ckm1}
 \end{eqnarray}
As reported in Ref.~\cite{ckmfitter} the best-fit values of the parameters $\lambda$, $A$, $\bar{\rho}$,
$\bar{\eta}$ with $1\sigma$ errors are
 \begin{eqnarray}
  \lambda &=& \sin\theta_{C}=0.22543\pm0.00077~,~~~~~A=0.812^{+0.013}_{-0.027}~, \nonumber\\
  \bar{\rho} &=& 0.144\pm0.025~,~~~~~~~~~~~~~~~~~~~~~~~\bar{\eta}=0.342^{+0.016}_{-0.015}~,
 \end{eqnarray}
where $\bar{\rho}=\rho(1-\lambda^{2}/2)$ and $\bar{\eta}=\eta(1-\lambda^{2}/2)$. The effects caused by
CP violation are always proportional to the Jarlskog invariant~\cite{Jarlskog:1985ht}, defined as
$J^{\rm quark}_{CP} = -{\rm Im}[V_{ud}V_{cs}V^{\ast}_{us}V^{\ast}_{cd}] \simeq A^{2} \lambda^{6}\eta$
whose value is $2.96^{+0.18}_{-0.17}\times10^{-5}$ at $1\sigma$ level~\cite{ckmfitter}.  From
Eqs.~(\ref{CHcorrect}), (\ref{LL}) and (\ref{mixing}), the PMNS matrix in Eq.~(\ref{mixing relation})
can be expressed as
 \begin{eqnarray}
 \small
  U_{\rm PMNS}=
   {\left(\begin{array}{ccc}
  V^{\ell}_{L11}V_{11}-V^{\ell}_{L12}V_{21}+V^{\ell\ast}_{L31}V_{31}
   & \frac{V^{\ell}_{L11}-V^{\ell}_{L12}+V^{\ell\ast}_{L31}}{\sqrt{3}}
   & V^{\ell}_{L11}V_{13}-V^{\ell}_{L12}V_{23}+V^{\ell\ast}_{L31}V_{33} \\
  V^{\ell}_{L22}V_{21}-V^{\ell}_{L23}V_{31}+V^{\ell\ast}_{L12}V_{11}
   & \frac{V^{\ell}_{L22}-V^{\ell}_{L23}+V^{\ell\ast}_{L12}}{\sqrt{3}}
   &  V^{\ell}_{L22}V_{23}-V^{\ell}_{L23}V_{33}+V^{\ell\ast}_{L12}V_{13} \\
  V^{\ell}_{L31}V_{33}+V^{\ell\ast}_{L13}V_{11}+V^{\ell\ast}_{L23}V_{21}
   & \frac{V^{\ell}_{L33}+V^{\ell\ast}_{L13}+V^{\ell\ast}_{L23}}{\sqrt{3}}
   & V^{\ell}_{L33}V_{33}+V^{\ell\ast}_{L13}V_{13}+V^{\ell\ast}_{L23}V_{23}
 \end{array}\right)} ,
  \label{PMNS2}
 \end{eqnarray}
where $V^{\ell}_{Lij}$ is the $(ij)$-element of the mixing matrix $V^{\ell}_L$, and $V_{ij}$ is the
$(ij)$-element of $U^{\dag}_{\omega}U_{\nu}$ given by
 \begin{eqnarray}
 V = U^{\dag}_{\omega}U_{\nu}=e^{i\pi/2}{\left(\begin{array}{ccc}
 \frac{ce^{i\varphi_{1}}+se^{i\varphi_{2}}}{\sqrt{3}} & \frac{1}{\sqrt{3}}
  &  \frac{ce^{i\varphi_{2}}-se^{i\varphi_{1}}}{\sqrt{3}} \\
 -\frac{ce^{i(\varphi_{1}+\frac{\pi}{3})}+se^{i(\varphi_{2}-\frac{\pi}{3})}}{\sqrt{3}}
  &  \frac{1}{\sqrt{3}}
  &  \frac{se^{i(\varphi_{1}+\frac{\pi}{3})}-ce^{i(\varphi_{2}-\frac{\pi}{3})}}{\sqrt{3}} \\
 -\frac{ce^{i(\varphi_{1}-\frac{\pi}{3})}+se^{i(\varphi_{2}+\frac{\pi}{3})}}{\sqrt{3}}
  &  \frac{1}{\sqrt{3}}
  &  \frac{se^{i(\varphi_{1}-\frac{\pi}{3})}-ce^{i(\varphi_{2}+\frac{\pi}{3})}}{\sqrt{3}}
 \end{array}\right)}~.
 \label{PMNS1}
 \end{eqnarray}
Here $s\equiv\sin\theta$ and $c\equiv\cos\theta$, and the common phase has no physical meaning so that
it can be neglected. By recasting Eq.~(\ref{PMNS2}) with the transformations $e \to e ~e^{i\alpha_{1}}$,
$\mu \to \mu ~e^{i\beta_{1}}$, $\tau \to \tau ~e^{i\beta_{2}}$ and
$\nu_{2} \to \nu_{2} ~e^{i(\alpha_{1}-\alpha_{2})}$, we can rewrite the PMNS matrix as
 \begin{eqnarray}
  U_{\rm PMNS}=
 {\left(\begin{array}{ccc}
 |U_{e1}| & |U_{e2}| & U_{e3}e^{-i\alpha_{1}} \\
 U_{\mu1}e^{-i\beta_{1}} & U_{\mu2}e^{i(\alpha_{1}-\alpha_{2}-\beta_{1})} &  |U_{\mu3}| \\
 U_{\tau1}e^{-i\beta_{2}} & U_{\tau2}e^{i(\alpha_{1}-\alpha_{2}-\beta_{2})} & |U_{\tau3}|
 \end{array}\right)}~
 \label{PMNS}
 \end{eqnarray}
which corresponds to the standard parametrization as in PDG~\cite{pdg}.
From the above equation, the neutrino mixing parameters can be displayed as
 \begin{eqnarray}
  \sin^{2}\theta_{12}&=&\frac{|U_{e2}|^{2}}{1-|U_{e3}|^{2}}~,~~~
   \sin^{2}\theta_{23}=\frac{|U_{\mu3}|^{2}}{1-|U_{e3}|^{2}}~,\nonumber\\
  \sin^{2}\theta_{13}&=&|U_{e3}|^{2}~,~~~~~~~~~~
   \delta_{CP}=\alpha_{1}-\alpha_{3}~,
 \label{mixing1}
 \end{eqnarray}
where $\alpha_{1} = \arg(U_{e1})$, $\alpha_{2} = \arg(U_{e2})$, $\alpha_{3} = \arg(U_{e3})$,
$\beta_{1} = \arg(U_{\mu3})$ and $\beta_{2} = \arg(U_{\tau3})$.
Leptonic CP violation at low energies can be detected through the neutrino oscillations which are
sensitive to the Dirac CP-phase, but insensitive to the Majorana CP-phases in
$U_{\rm PMNS}$~\cite{Branco:2002xf}:
the Jarlskog invariant $J_{CP}\equiv{\rm Im}[U_{e1}U_{\mu2}U^{\ast}_{e2}U^{\ast}_{\mu1}]$, where
$U_{\alpha j}$ is an element of the PMNS matrix in Eq.~(\ref{PMNS2}), with $\alpha=e,\mu,\tau$
corresponding to the lepton flavors and $j=1,2,3$ corresponding to the light neutrino mass eigenstates.
To see how both CP phases $\varphi_{21}$ (coming from the neutrino sector) and $\phi^{\ell}_{1,2,3}$
(coming from the charged lepton sector) are correlated with low energy CP violation measurable through
the neutrino oscillations, let us consider the leptonic CP violation parameter $J_{CP}$. Corresponding
to the three different types of $V^{\ell}_L$ given in Eqs.~(\ref{LL}), (\ref{LL1}) and (\ref{LL2}),
three different $J_{CP}$ are displayed as $J^{\rm I}_{CP}$, $J^{\rm II}_{CP}$ and $J^{\rm III}_{CP}$,
respectively:
 \begin{eqnarray}
  J^{\rm I}_{CP}&\simeq&\frac{\cos2\theta}{6\sqrt{3}}
   +\frac{\lambda\sqrt{3}}{9}\sin2\theta\sin\phi^{\ell}_{3}\cos(\pi/6+\varphi_{21}) \nonumber\\
  &-&\frac{\lambda^{2}}{3\sqrt{3}}\Big( \cos2\theta
   -B_{1}\sin2\theta\sin\phi^{\ell}_{1}\sin\varphi_{21} \Big) +{\cal O}(\lambda^{3})~,\label{JCP1}\\
  J^{\rm II}_{CP}&\simeq&\frac{\cos2\theta}{6\sqrt{3}}
   +\frac{\lambda\sqrt{3}}{9}\sin2\theta\sin\phi^{\ell}_{3}\cos(\pi/6+\varphi_{21})\nonumber\\
  &-&\frac{\lambda^{2}}{9}\Big\{\sqrt{3}\cos2\theta
   -A_{2} \Big[ \sqrt{3}\cos2\theta\cos\phi^{\ell}_{2}
   +\sin\phi^{\ell}_{2}\big(1-\sin2\theta\cos(\pi/3+\varphi_{21}) \big) \Big] \nonumber\\
  &-&\sqrt{3}B_{1}\sin2\theta\sin\phi^{\ell}_{1}\sin\varphi_{21}\Big\}
   +{\cal O}(\lambda^{3}) ~, \label{JCP2}\\
  J^{\rm III}_{CP}&\simeq&\frac{\cos2\theta}{6\sqrt{3}}
   +\frac{\lambda\sqrt{3}}{9}\sin2\theta\sin\phi^{\ell}_{3}\cos(\pi/6+\varphi_{21})
  -\frac{\lambda^{2}}{9}\Big\{\sqrt{3}\cos2\theta\nonumber\\
  &-& A_{3} \Big[ \sqrt{3}\cos2\theta\cos\phi^{\ell}_{2}+\sin\phi^{\ell}_{2}
   -\sin2\theta\sin\phi^{\ell}_{2}\cos(\pi/3+\varphi_{21}) \Big] \Big\} +{\cal O}(\lambda^{3})~.
 \label{JCP3}
 \end{eqnarray}
Note that $J^{\rm I}_{CP}$, $J^{\rm II}_{CP}$ and $J^{\rm III}_{CP}$ have exactly the same expressions,
up to ${\cal O}(\lambda)$.  Also, ~$\sin\phi^{\ell}_{3}$~ appears commonly in the terms of the first
order in $\lambda$, and its value is crucial to satisfy the neutrino data for the solar mixing angle
and the first QLC relation given in Eq.~(\ref{QLC}).
In particular, for $\theta \to \pi/4 ~({\rm or}~ -\pi/4)$ and $\varphi_{21} \to 0 ~({\rm or}~ \pi)$,
we obtain $J^{\rm I}_{CP} \simeq J^{\rm II}_{CP} \simeq J^{\rm III}_{CP} \simeq \pm \lambda/6$
for $\sin\phi^{\ell}_{3} \simeq \pm 1$.

\section{The QLC relations and the charged lepton mixing}

In this section we investigate the possibility that in our framework based on the discrete $A_4$ flavor
symmetry, the QLC relations hold in a natural way through the mixing matrices obtained in the previous
section.
Due to the form of the PMNS matrix $U_{\rm PMNS} = V^{\ell\dag}_{L} U^{\dag}_{\omega} U_{\nu}$, the
CKM-like mixing matrix $V^{\ell}_{L}$ induced from the charged lepton sector becomes a key ingredient
for this purpose.  In particular, as we shall see, certain {\it phases} of the elements of
$V^{\ell}_{L}$ plays an important role to satisfy the QLC relations.

For our numerical analysis, we use the five neutrino experimental data of
$\Delta m_{\rm sol}^2$, $\Delta m_{\rm atm}^2$, $\theta_{12}$, $\theta_{13}$ and $\theta_{23}$ at
$3\sigma$ level given in Table~\ref{tab:data}~\cite{GonzalezGarcia:2010er} as inputs.
For our purpose, we will consider only the normal hierarchical mass ordering case of the light neutrinos,
where $m_0$ is taken as order $10^{-2}$ eV.  By using the relation $m_{0}=x^{2}\upsilon^{2}/M$ with
the seesaw scale $M = 10^{12}$ GeV and the SM Higgs VEV $\upsilon = 174$ GeV, the values of the relevant
parameters are taken as
 \begin{eqnarray}
  &&
  0.5 < \kappa < 1.5 ~,~~~ 0.01 < x < 0.02 ~,~~~
  0.0001 < y_{1,2} < 0.1 ~, \nonumber\\
  && 0 \leq \phi_{1,2,3}^{\ell} \leq 2\pi~,~~~  0 \leq \rho_{1,2} \leq 2\pi ~.
  \label{input}
 \end{eqnarray}
Without loss of generality, we take $A_{1,2,3} = B_{1,2} = 1$ appearing in the charged lepton mixing
matrix $V^{\ell}_{L}$, because they do not affect the leptonic mixing parameters significantly.
From now on, we will discuss three different scenarios corresponding to the three different forms of
$V^{\ell}_{L}$ obtained in Eqs.~(\ref{LL}), (\ref{LL1}) and (\ref{LL2}).

\subsection{Scenario-I}

\begin{figure}[t]
 \begin{minipage}[t]{6.0cm}
  \epsfig{figure=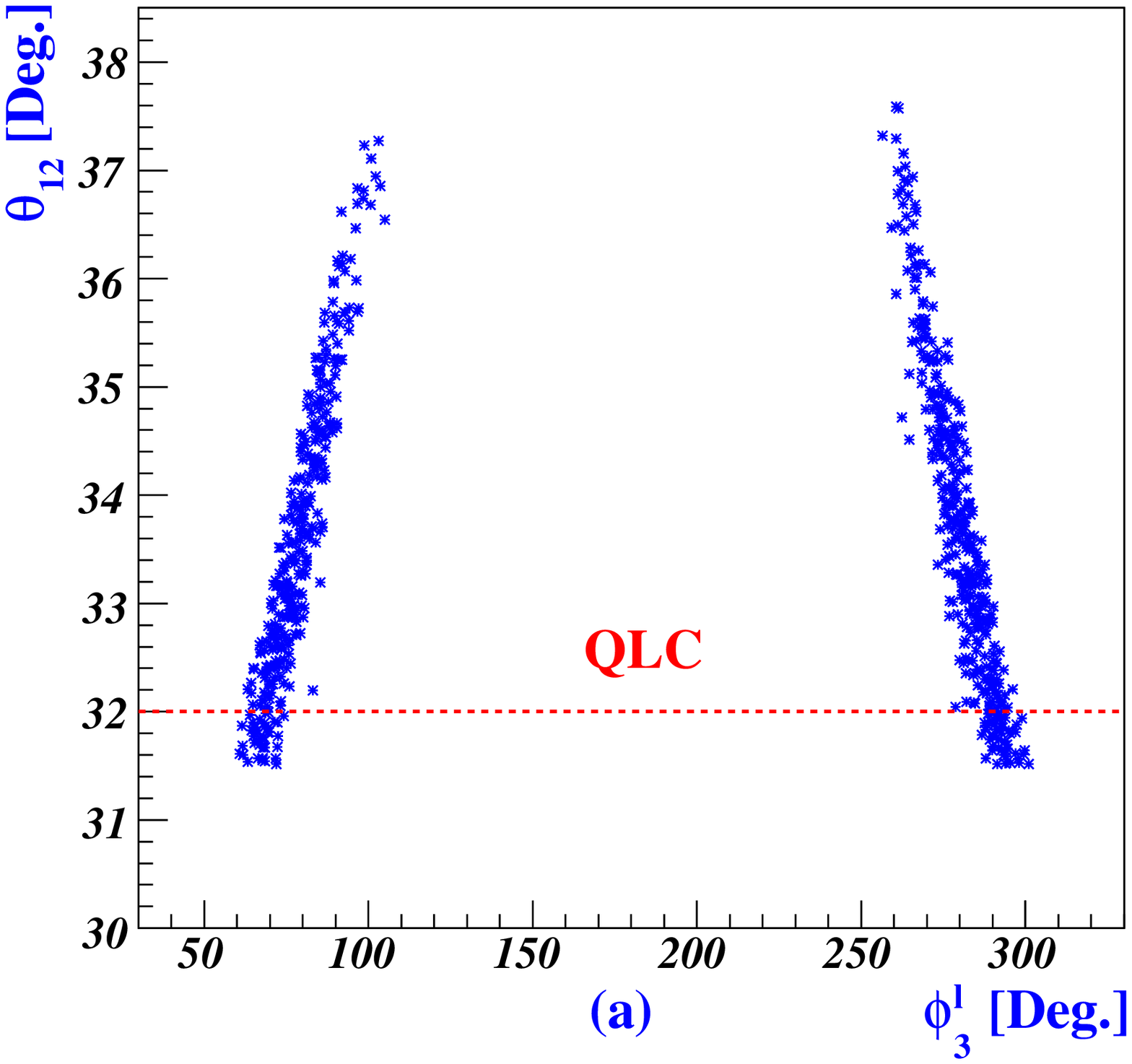,width=6.5cm,angle=0}
 \end{minipage}
 \hspace*{1.0cm}
 \begin{minipage}[t]{6.0cm}
  \epsfig{figure=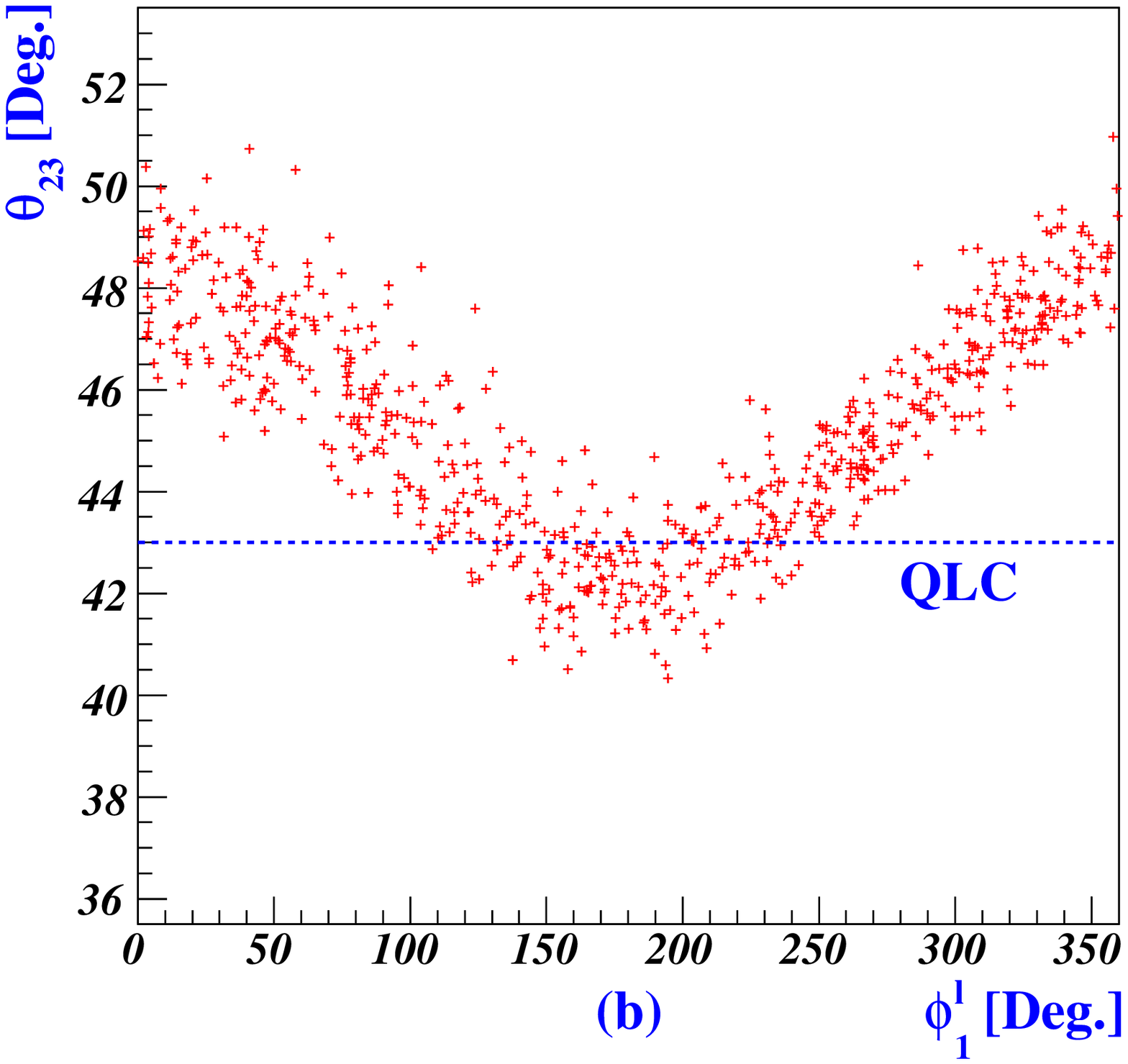,width=6.5cm,angle=0}
 \end{minipage}
 \caption{\label{FigA1}
  Plots in Scenario-I displaying (a) the allowed regions of the solar mixing angle $\theta_{12}$
  versus the CP phase $\phi^{\ell}_{3}$, and (b) the allowed regions of the atmospheric
  mixing angle $\theta_{23}$ versus the CP phase $\phi^{\ell}_{1}$.
  The horizontal dotted lines in both plots correspond to $\theta_{12} = 32^{\circ}$ and
  $\theta_{23} = 43^{\circ}$, respectively, satisfying the exact QLC relations with
  $\theta^{q}_{12}=13^{\circ}$ and $\theta^{q}_{23}=2^{\circ}$.}
\end{figure}

Let us discuss the first scenario in which the charged lepton mixing matrix $V^{\ell}_{L}$ is the same
in form as the CKM matrix except for the different phases of each matrix element, as given in
Eq.~(\ref{LL}).
From the form of $U_{\rm PMNS}$ given in Eq.~(\ref{PMNS}), the solar neutrino mixing angle $\theta_{12}$
can be approximated, up to order $\lambda^{3}$, as
 \begin{eqnarray}
  \sin^{2}\theta_{12} &=& \frac{1-2\lambda\cos\phi^{\ell}_{3}
   +\lambda^{3}(\cos\phi^{\ell}_{3}-2A_{1}\cos\phi^{\ell}_{2}
   +2B_{1}\cos\tilde{\phi}^{\ell}_{13})}
   {2+\sin2\theta\cos\varphi_{21}-\Xi\lambda-\lambda^{2}\Psi_{1}-\Omega_{1}\lambda^{3}}~,
  \label{angle12I}
 \end{eqnarray}
where $\tilde{\phi}^{\ell}_{ij}\equiv\phi^{\ell}_{i}+\phi^{\ell}_{j}$ and the parameters $\Psi_{1}$
and $\Omega_{1}$ are defined as
 \begin{eqnarray}
  \Psi_{1}&=&\sqrt{3}\sin2\theta\cos(\varphi_{21}+\pi/6)~,\nonumber\\
  \Omega_{1}&=&\Theta_{1} +A_{2} \Big[ \sqrt{3}\cos2\theta\sin\phi^{\ell}_{2}
   -\cos\phi^{\ell}_{2}(1-2\sin2\theta\cos(\varphi_{21}+\pi/3)) \Big] ~,
 \label{12parameterI}
 \end{eqnarray}
with
 \begin{eqnarray}
 \Theta_{1}= \frac{\Xi}{2}
  +B_{2} \Big[ \cos\tilde{\phi}^{\ell}_{13}(1-2\sin2\theta\cos(\varphi_{21}+\pi/3))
   -\sqrt{3}\cos2\theta\sin\tilde{\phi}^{\ell}_{13} \Big]~.
 \end{eqnarray}
The parameter $\Xi$ defined as
 \begin{eqnarray}
  \Xi =\cos\phi^{\ell}_{3}+\sqrt{3}\sin\phi^{\ell}_{3}\cos2\theta
   -\sin2\theta \Big[ \cos\big(\varphi_{21}-\phi^{\ell}_{3}-\frac{\pi}{3}\big)
   +\cos\big(\varphi_{21}+\phi^{\ell}_{3}-\frac{\pi}{3}\big) \Big] ~,
  \label{Xi}
 \end{eqnarray}
appears in all the three scenarios, as we shall see.
In Eq.~(\ref{angle12I}), if we turn off the higher dimensional operators in the Lagrangian,
that is, if $\theta \to \pm \pi/4$, $\varphi_{21} \to 0(\pi)$ and $\lambda \to 0$, the
TBM angle $\sin^{2}\theta_{12}=1/3$ is restored, as expected.
In the limit of $\theta\rightarrow\pi/4$ and $\varphi_{21}\rightarrow0$ (inverted hierarchy of the
neutrino masses), or $\theta\rightarrow-\pi/4$ and $\varphi_{21}\rightarrow\pi$ (normal hierarchy of the
neutrino masses), the parameters behave as ~$\Xi\rightarrow 0$, $\Psi_{1}\rightarrow 3/2$ and
$\Omega_{1}\rightarrow 0$.
From Eq.~(\ref{angle12I}) we see that the solar mixing angle $\theta_{12}$ depends strongly on the CP
phase $\phi^{\ell}_{3}$ which comes from the elements of the charged lepton mixing matrix $V^{\ell}_L$.
Since there is no $\lambda^{2}$ term in the numerator of Eq.~(\ref{angle12I}), the allowed values of the
phase $\phi^{\ell}_{3}$ are within the narrower range, compared to those of the other two scenarios
[see Eqs.~(\ref{angle12II}) and (\ref{angle12III})].
The left plot of Fig.~\ref{FigA1} shows that the first QLC relation in Eq.~(\ref{QLC}) can be satisfied
for the values of $\phi^{\ell}_{3}$ in the range of $0.259\lesssim\cos\phi^{\ell}_{3} \lesssim 0.423$.
Thus, in this scenario the phase term $\cos \phi^{\ell}_3$ originating from the
dimension-5 operators plays a key role in explaining the first QLC relation
$\theta_{12}+\theta^{q}_{12} = \pi/4$.

The atmospheric mixing angle $\theta_{23}$ can be approximated, up to order $\lambda^{3}$, as
 \begin{eqnarray}
  \sin^{2}\theta_{23}=\frac{1-\sin2\theta\cos(2\pi/3-\varphi_{21})
   -\Xi\lambda-\lambda^{2}\Upsilon_{1}+\lambda^{3}\Theta_{1}}{2+\sin2\theta\cos\varphi_{21}
   -\Xi\lambda-\lambda^{2}\Psi_{1}-\lambda^{3}\Omega_{1}} ~,
  \label{angle23I}
 \end{eqnarray}
where the parameter $\Upsilon_{1}$ is defined as
 \begin{eqnarray}
  \Upsilon_{1}=\Psi_{1}-B_{1} \Big[ \cos\phi^{\ell}_{1}(1+2\cos\varphi_{21}\sin2\theta)
   +\sqrt{3}\sin\phi^{\ell}_{1}\cos2\theta \Big] ~,
 \label{23parameterI}
 \end{eqnarray}
If the contributions from the higher dimensional operators are neglected, that is, if
$\theta \to \pm \pi/4$, $\varphi_{21} \to 0(\pi)$ and $\lambda \to 0$, the TBM angle
$\sin^{2}\theta_{23}=1/2$ is restored, as expected.
In the limit of $\theta \to \pm \pi/4$ and $\varphi_{21} \to 0(\pi)$, the parameters
$\Upsilon_{1}$ and $\Theta_{1}$ behave as $\Upsilon_{1}\rightarrow3/2-3B_{1}\cos\phi^{\ell}_{1}$ and
$\Theta_{1}\rightarrow0$.  In this limit, the atmospheric mixing angle becomes
 \begin{eqnarray}
  \sin^{2}\theta_{23}\approx \frac{1}{2}+\Big(\cos\phi^{\ell}_{1}-\frac{1}{4}\Big)\lambda^{2}~,
  \label{theta23AppI}
 \end{eqnarray}
where $B_{1}=1$ is used. The above equation shows that a deviation from the maximality of atmospheric
mixing angle depends mainly on the value of $\cos\phi^{\ell}_{1}$: the second QLC relation in
Eq.~(\ref{QLC}) can be satisfied if $\cos\phi^{\ell}_{1}\approx-0.43$.
The right plot of Fig.~\ref{FigA1} shows the behavior of the atmospheric mixing angle $\theta_{23}$
as a function of $\phi^{\ell}_{1}$. To satisfy the second QLC relation, the value of $\phi^{\ell}$
should be in the range of $90^{\circ}\lesssim\phi^{\ell}_{1}\lesssim270^{\circ}$.
Also, the lower bound on $\theta_{23}$ is obtained as $\theta_{23}\gtrsim41^{\circ}$ for
$\cos\phi^{\ell}_{1}\simeq-1$.
We note again that the effects of the dimension-5 operators, e.g., responsible for the phase term
$\cos\phi^{\ell}_1$, are the key ingredients for accommodating the QLC relations.

\begin{figure}[t]
 \begin{minipage}[t]{6.0cm}
  \epsfig{figure=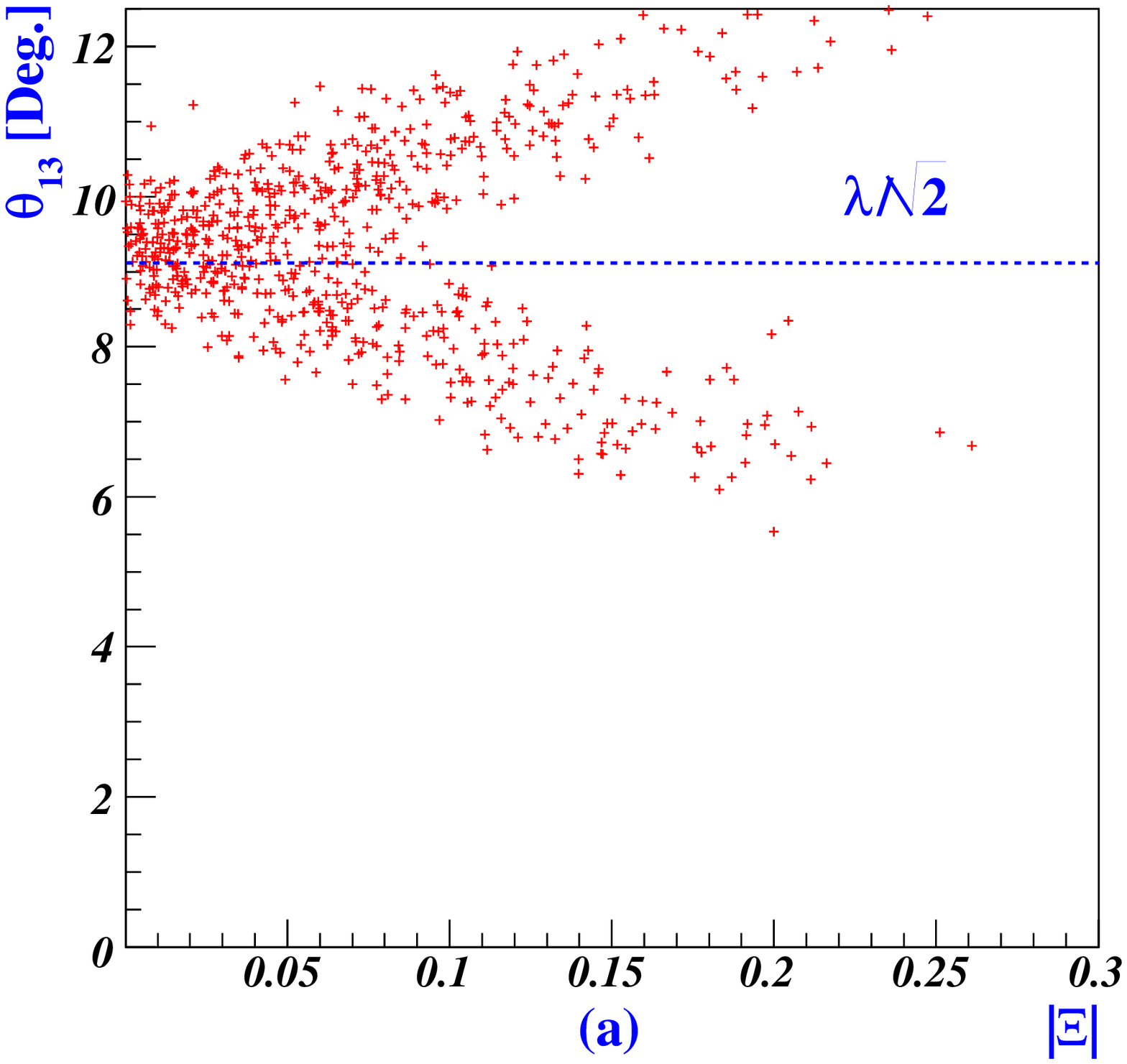,width=6.5cm,angle=0}
 \end{minipage}
 \hspace*{1.0cm}
 \begin{minipage}[t]{6.0cm}
  \epsfig{figure=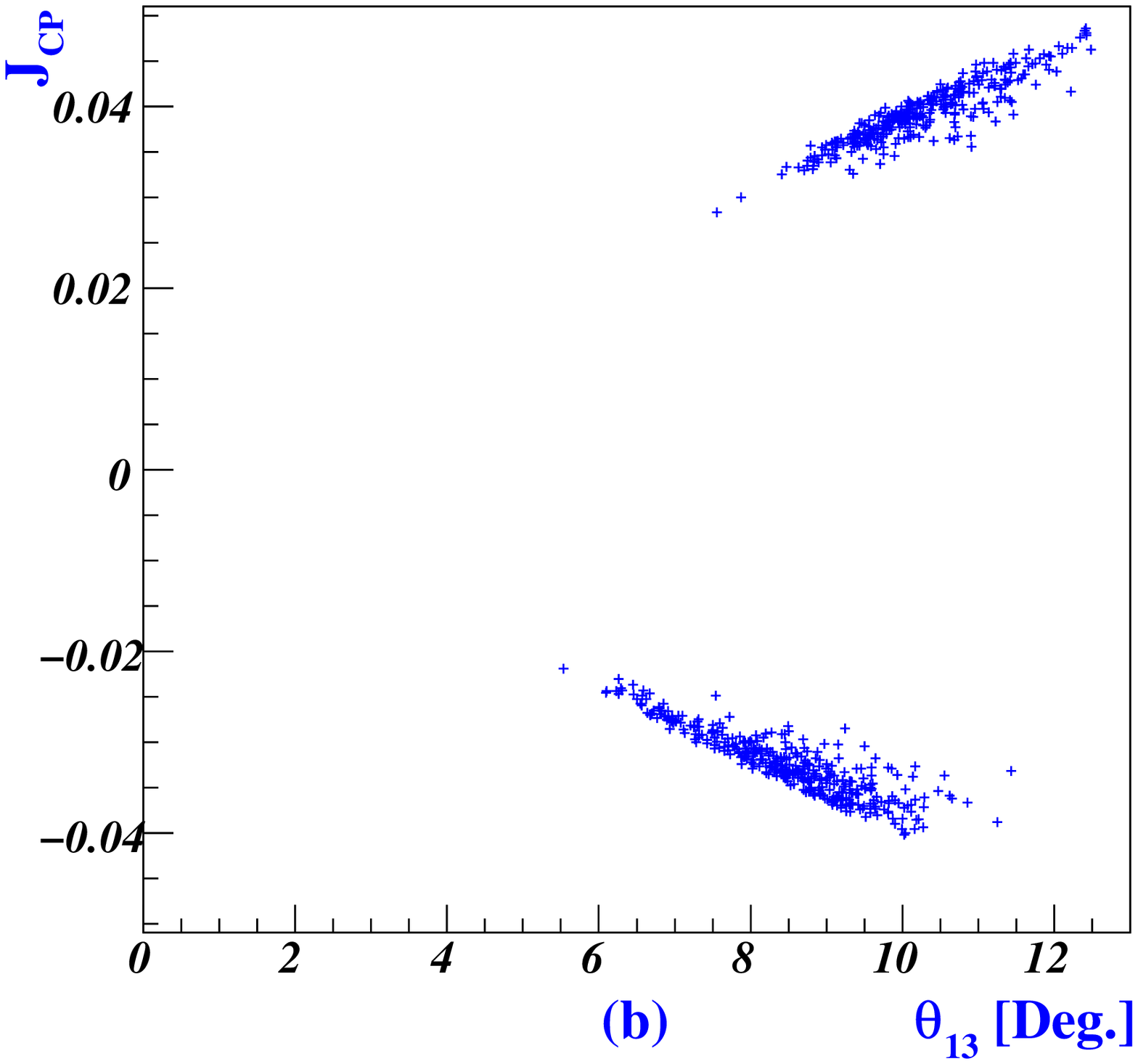,width=6.5cm,angle=0}
 \end{minipage}
 \caption{\label{FigA2}
  In Scenario-I, allowed values for (a) the reactor angle $\theta_{13}$  and
  (b) $J_{\rm CP}$ as a function of $|\Xi|$ and $\theta_{13}$, respectively.
  The horizontal dotted line in the left figure corresponds to $\theta_{13} = 9.1^{\circ}$.}
\end{figure}

The reactor angle $\theta_{13}$ can be expressed approximately as
 \begin{eqnarray}
  \sin\theta_{13}
  =\frac{1}{\sqrt{3}}\sqrt{1-\sin2\theta\cos\varphi_{21}
   +\Xi\lambda+\lambda^{2}\Psi_{1}+\lambda^{3}\Omega_{1}}~.
  \label{Theta13I}
 \end{eqnarray}
In the limit of $\theta \to \pm \pi/4$ and $\varphi_{21} \to 0(\pi)$, the parameters
$\Xi$, $\Psi_{1}$, $\Omega_{1}$ behave in the same way as before, which leads to
$\sin\theta_{13} \to \lambda/\sqrt{2}$~\cite{Chauhan:2006im}.
Then, the expression of the reactor angle can be simplified as
 \begin{eqnarray}
  \sin\theta_{13} \simeq \sqrt{\frac{\lambda^{2}}{2}
   +\frac{\varepsilon+\Xi\lambda}{3}}~,
 \label{sintheta13_1}
 \end{eqnarray}
with $\varepsilon=1-\sin2\theta\cos\varphi_{21}$.
Since $\varepsilon \geq 0$, depending on the sign of $\Xi$, the second term in the squared-root can
increase or decrease the value of $\sin \theta_{13}$ around the center value $\lambda / \sqrt{2}$.
Furthermore, since the value of $|\Xi|$ is bounded as can be seen in Eq.~(\ref{Xi}), we expect
that there is a lower bound on the possible value of $\theta_{13}$.
The parameter $\Xi$ depends mainly on $y_1$ and $y_2$, defined in Eq.~(\ref{yukawaNu}), which represent
the effects of the dimension-5 operators.
Thus, in this scenario the lower bound on the mixing angle $\theta_{13}$ is strongly dependent on
the cutoff scale $\Lambda$, the $A_{4}$ symmetry breaking scale $\upsilon_{\chi}$ and the relevant
couplings $|y^{s,a}_{N}|$, through $y_1$ and $y_2$.
For example, if one takes $\Lambda = 10^{15}$ GeV, $\upsilon_{\chi} = 10^{12}$ GeV,
$|y^{s,a}_{N}|\sim{\cal O}(1)$ together with $x\sim{\cal O}(0.01)$, then one obtains
$y_{1,2}\sim{\cal O}(0.1)$ which determines the lower bound on $\theta_{13}$.
The left plot of Fig.~\ref{FigA2} shows the behavior of $\theta_{13}$ as a function of $|\Xi|$, where
there is the lower bound $\theta_{13}\gtrsim5^{\circ}$ and the horizontal dotted line represents
$\theta_{13}=9.1^{\circ}$.
Since neutrino oscillation experiments are sensitive to the Dirac CP phase $\delta_{CP}$,
the Jarlskog invariant of the leptonic sector given in Eq.~(\ref{JCP1}) would be a signal of CP
violation.
The right plot of Fig.~\ref{FigA2} shows our prediction for the Jarlskog invariant
~$|J_{\rm CP}| \approx 0.02-0.05$~ due to the sizable $\theta_{13}$.
This can be tested in the future experiments such as the upcoming long baseline neutrino
oscillation ones.

\subsection{Scenario-II}

\begin{figure}[t]
 \begin{minipage}[t]{6.0cm}
  \epsfig{figure=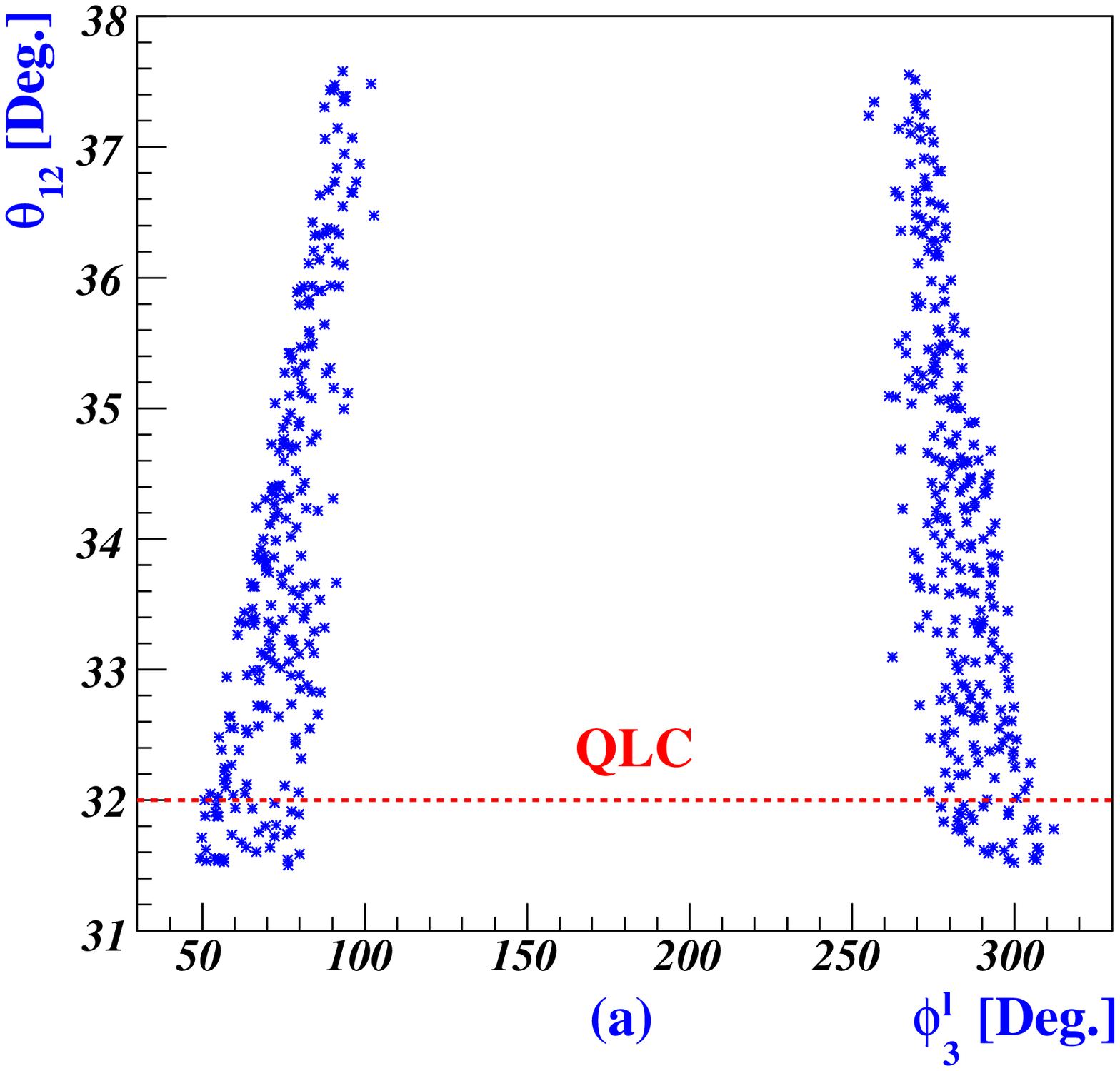,width=6.5cm,angle=0}
 \end{minipage}
 \hspace*{1.0cm}
 \begin{minipage}[t]{6.0cm}
  \epsfig{figure=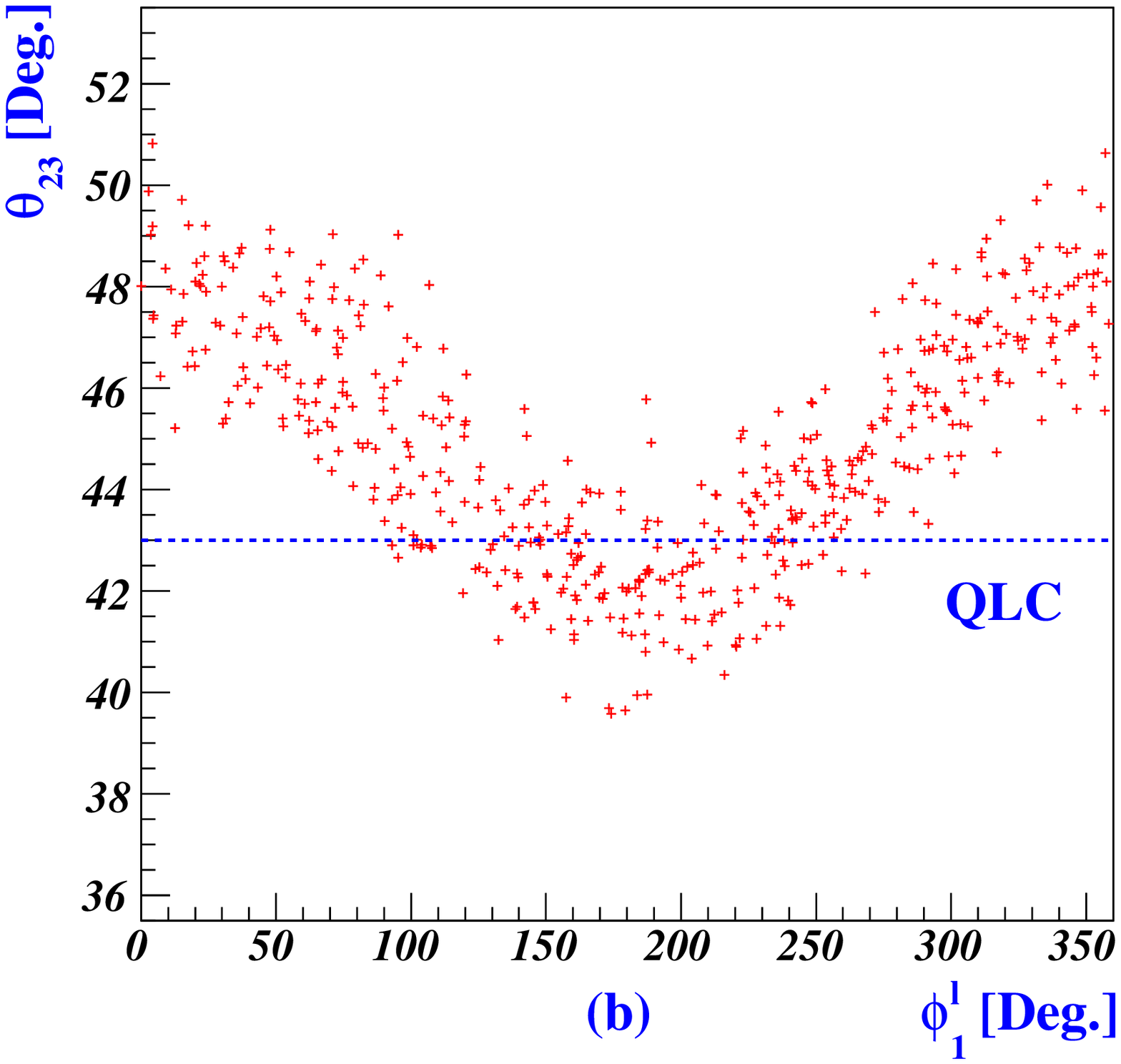,width=6.5cm,angle=0}
 \end{minipage}
 \caption{\label{FigB1}
  Same as Fig.~\ref{FigA1} except for Scenario-II. }
\end{figure}

Now we turn to the second scenario in which the charged leptonic mixing matrix $V^{\ell}_L$ is given
by Eq.~(\ref{LL1}).
Similarly to Scenario-I, from $U_{\rm PMNS}$ in Eq.~(\ref{PMNS}), the solar neutrino mixing angle
$\theta_{12}$ can be approximated, up to order $\lambda^{3}$, as
 \begin{eqnarray}
  \sin^{2}\theta_{12} &=& \frac{1-2\lambda\cos\phi^{\ell}_{3}
   -2A_{2}\lambda^{2}\cos\phi^{\ell}_{2}+\lambda^{3}(\cos\phi^{\ell}_{3}
   -2A_{2}\cos\phi^{\ell}_{32}+2B_{2}\cos\tilde{\phi}^{\ell}_{13})}
   {2+\sin2\theta\cos\varphi_{21}-\Xi\lambda-\lambda^{2}\Psi_{2}-\Omega_{2}\lambda^{3}}~,
  \label{angle12II}
 \end{eqnarray}
where $\phi^{\ell}_{ij}\equiv\phi^{\ell}_{i}-\phi^{\ell}_{j}$, and
 \begin{eqnarray}
  \Psi_{2} &=& \sqrt{3}\sin2\theta\cos(\varphi_{21}+\pi/6)\nonumber\\
  &-& A_{2} \Big[ \sqrt{3}\cos2\theta\sin\phi^{\ell}_{2}-\cos\phi^{\ell}_{2}
   -2\cos\phi^{\ell}_{2}\sin2\theta\cos(\varphi_{21}+\pi/3) \Big] ~,\nonumber\\
  \Omega_{2} &=& \Theta_{2} +A_{2} \Big[ (1+2\sin2\theta\cos\varphi_{21})\cos\phi^{\ell}_{32}
   -\sqrt{3}\cos2\theta\sin\phi^{\ell}_{32} \Big] ~,
  \label{12parameterII}
 \end{eqnarray}
with
 \begin{eqnarray}
 \Theta_{2}= \frac{\Xi}{2}+B_{2} \Big[ \cos\tilde{\phi}^{\ell}_{13}
  (1 -2\sin2\theta\cos(\varphi_{21}+\pi/3))
  -\sqrt{3}\cos2\theta\sin\tilde{\phi}^{\ell}_{13} \Big] ~.
 \end{eqnarray}
From Eq.~(\ref{angle12II}), neglecting the contributions from the higher dimensional operators,
that is, for $\theta \to \pm \pi/4$, $\varphi_{21} \to 0(\pi)$ and $\lambda \to 0$,
we see that the TBM angle $\sin^{2}\theta_{12}=1/3$ is restored, as expected.
In the limit of $\theta\rightarrow\pi/4$ and $\varphi_{21}\rightarrow0$ (inverted mass hierarchy of the
neutrinos), or $\theta\rightarrow-\pi/4$ and $\varphi_{21}\rightarrow\pi$ (normal mass hierarchy of the
neutrinos), the behaviors of the parameters $\Xi$, $\Psi_2$ and $\Omega_2$ are found to be:
$\Xi\rightarrow0$, $\Psi_{2}\rightarrow3/2$ and $\Omega_{2}\rightarrow3A_{2}\cos\phi^{\ell}_{32}$.
As in the case of Scenario-I, Eq.~(\ref{angle12II}) shows that the solar mixing angle depends strongly
on the phase $\phi^{\ell}_{3}$ arising from the elements of $V^{\ell}_L$.
The dependence of the solar mixing angle $\theta_{12}$ on the phase $\phi^{\ell}_3$ is plotted in the
left panel of Fig.~\ref{FigB1}, which shows that the first QLC relation,
$\theta_{12}+\theta^{q}_{12} = \pi/4$, can be satisfied provided that the value of $\phi^{\ell}_{3}$ is
in the range of $0.174\lesssim\cos\phi^{\ell}_{3}\lesssim0.643$.

\begin{figure}[t]
 \begin{minipage}[t]{6.0cm}
  \epsfig{figure=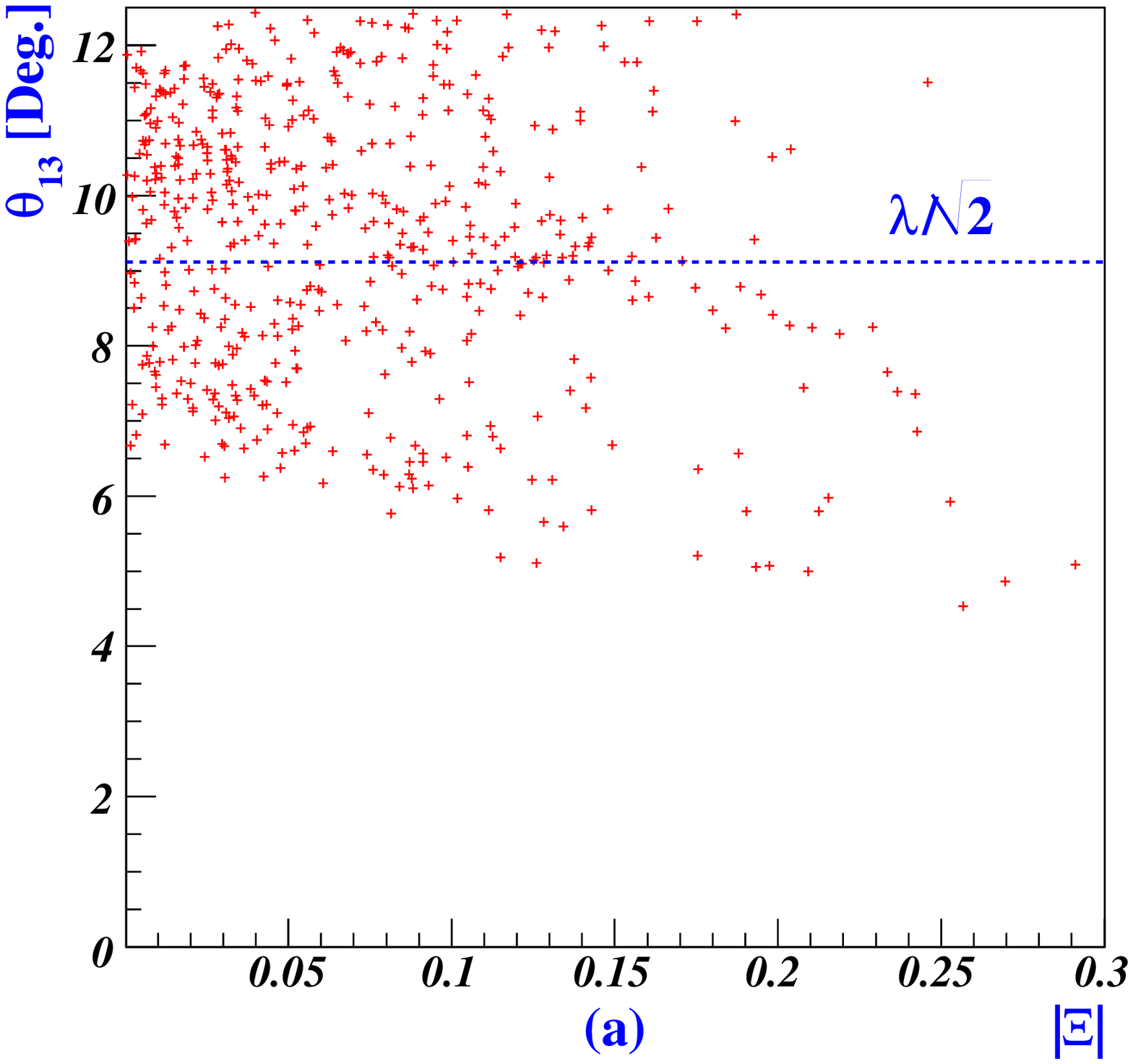,width=6.5cm,angle=0}
 \end{minipage}
 \hspace*{1.0cm}
 \begin{minipage}[t]{6.0cm}
  \epsfig{figure=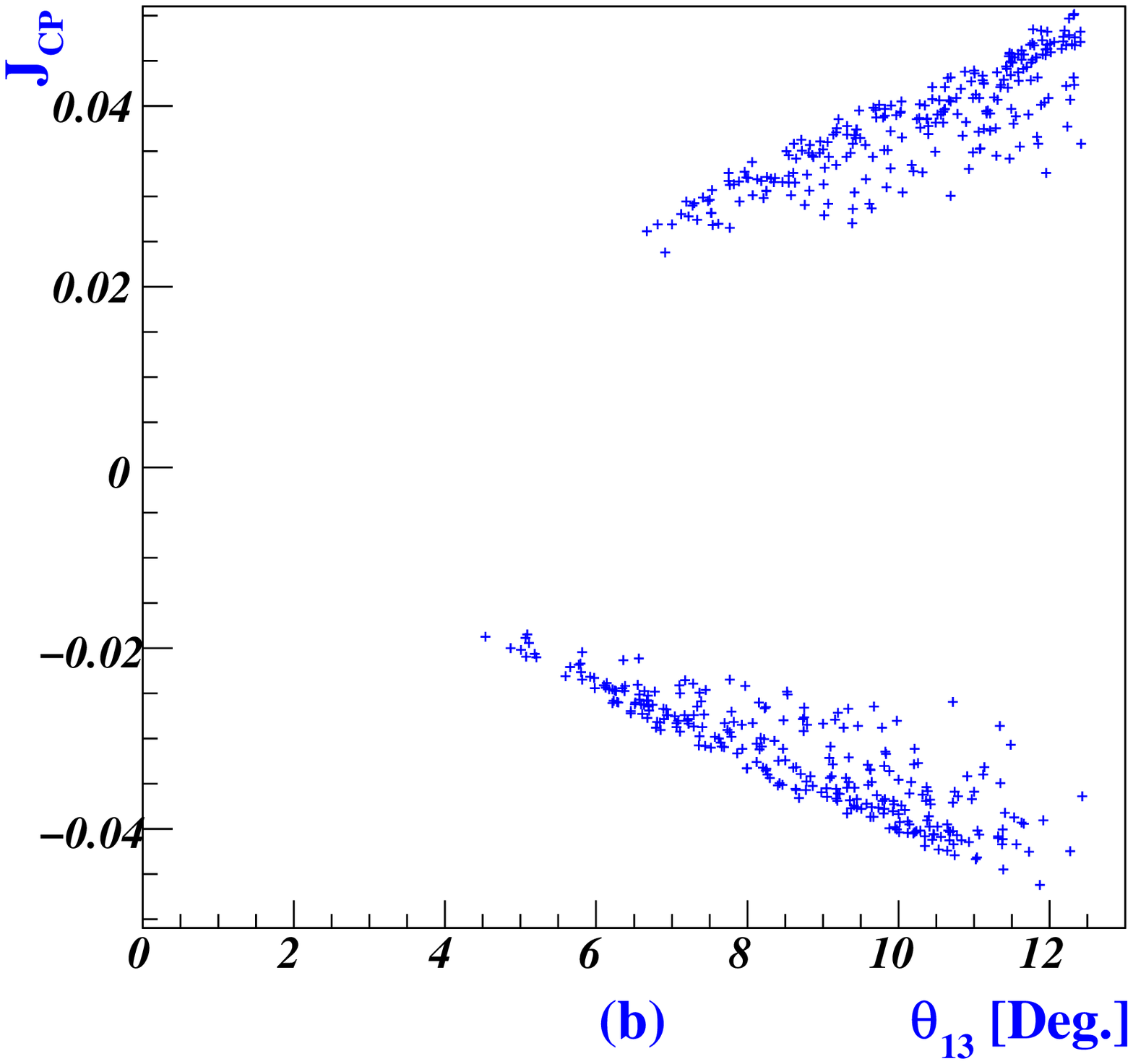,width=6.5cm,angle=0}
 \end{minipage}
 \caption{\label{FigB2}
  Same as Fig.~\ref{FigA2} except for Scenario-II. }
\end{figure}

The atmospheric mixing angle $\theta_{23}$ can be approximated as
 \begin{eqnarray}
  \sin^{2}\theta_{23}=\frac{1-\sin2\theta\cos(2\pi/3-\varphi_{21})
   -\Xi\lambda-\lambda^{2}\Upsilon_{2}+\lambda^{3}\Theta_{2}}
   {2+\sin2\theta\cos\varphi_{21} -\Xi\lambda-\lambda^{2}\Psi_{2}-\lambda^{3}\Omega_{2}} ~,
  \label{angle23II}
 \end{eqnarray}
where the parameter $\Upsilon_{2}$ is given by
 \begin{eqnarray}
  \Upsilon_{2}=\sqrt{3}\sin2\theta\cos(\varphi_{21}+\pi/6)
   -B_{2} \Big[ \cos\phi^{\ell}_{1}(1+2\cos\varphi_{21}\sin2\theta)
   +\sqrt{3}\sin\phi^{\ell}_{1}\cos2\theta \Big] ~.
\label{23parameterII}
 \end{eqnarray}
In the limit of $\theta \to \pm \pi/4$ and $\varphi_{21} \to 0(\pi)$, the parameters
$\Upsilon_2$ and $\Theta_2$ behaves as $\Upsilon_{2}\rightarrow3/2-3B_{2}\cos\phi^{\ell}_{1}$ and
$\Theta_{2}\rightarrow0$. Then, the atmospheric mixing angle can be expressed, up to order
$\lambda^{2}$, as
 \begin{eqnarray}
  \sin^{2}\theta_{23}\approx \frac{1}{2}+\Big(\cos\phi^{\ell}_{1}-\frac{1}{4}\Big)\lambda^{2}~,
  \label{theta23AppII}
 \end{eqnarray}
where $B_{2} = 1$ is used.
We note that the above equation is the same as Eq.~(\ref{theta23AppI}). Thus, in this limit, the
results obtained for Eq.~(\ref{theta23AppI}) in Scenario-I are also valid in this scenario.
The right plot of Fig.~\ref{FigB1} displays the dependence of $\theta_{23}$ on the phase
$\phi^{\ell}_{1}$ which leads to the conclusion that the values of $\phi^{\ell}_{1}$ satisfying the
second QLC relation, $\theta_{23}+\theta^{q}_{23} = \pi/4$, are in the range of
$90^{\circ}\lesssim\phi^{\ell}_{1}\lesssim270^{\circ}$.

The reactor angle $\theta_{13}$ can be written approximately as
 \begin{eqnarray}
  \sin\theta_{13}
  =\frac{1}{\sqrt{3}}\sqrt{1-\sin2\theta\cos\varphi_{21}
   +\Xi\lambda+\lambda^{2}\Psi_{2}+\lambda^{3}\Omega_{2}}~,
  \label{Theta13II}
 \end{eqnarray}
In the limit of $\theta \to \pm \pi/4$ and $\varphi_{21} \to 0(\pi)$, the parameters
$\Xi, \Psi_2, \Omega_2$ behave in the same way as before so that $\sin\theta_{13} \to
\lambda/\sqrt{2}$~\cite{Chauhan:2006im}. The expression of $\sin \theta_{13}$ can be simplified as
 \begin{eqnarray}
  \sin\theta_{13} \simeq \sqrt{\frac{\lambda^{2}}{2}
   +\frac{\varepsilon+\Xi\lambda}{3}}~.
 \label{sintheta13_2}
 \end{eqnarray}
This equation is the same as Eq.~(\ref{sintheta13_1}) in Scenario-I.
The left plot of Fig.~\ref{FigB2} shows how the mixing angle $\theta_{13}$ is constrained by the values
of the parameter $\Xi$.  There is a lower bound on the reactor angle: $\theta_{13}\gtrsim3.5^{\circ}$,
which is somewhat smaller than that found in Scenario-I.  In the right plot of Fig.~\ref{FigB2},
our prediction for the Jarlskog invariant given in Eq.~(\ref{JCP2}) is found to be
$|J_{\rm CP}| \approx 0.015-0.05$.

\subsection{Scenario-III}

\begin{figure}[t]
 \begin{minipage}[t]{6.0cm}
  \epsfig{figure=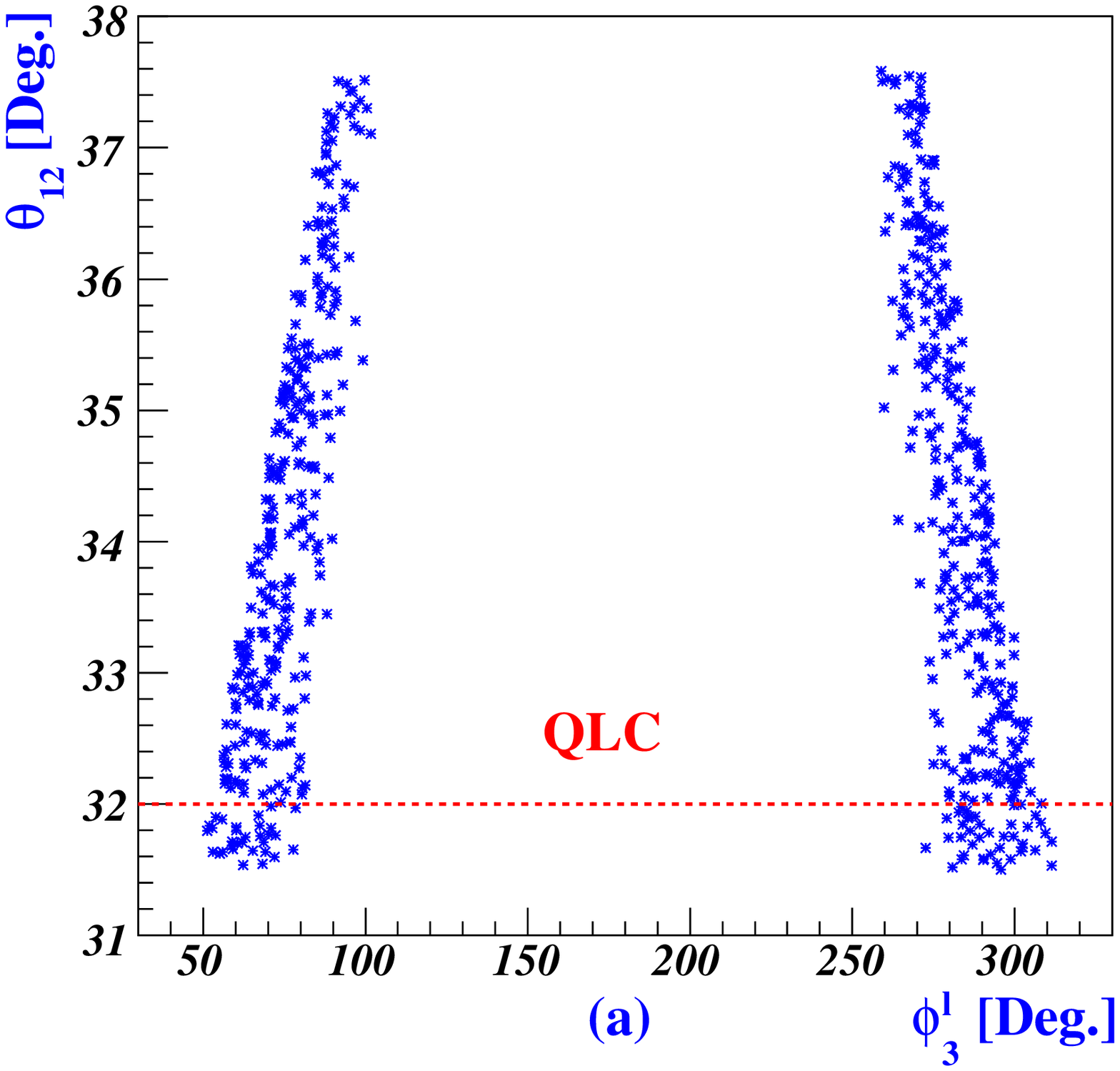,width=6.5cm,angle=0}
 \end{minipage}
 \hspace*{1.0cm}
 \begin{minipage}[t]{6.0cm}
  \epsfig{figure=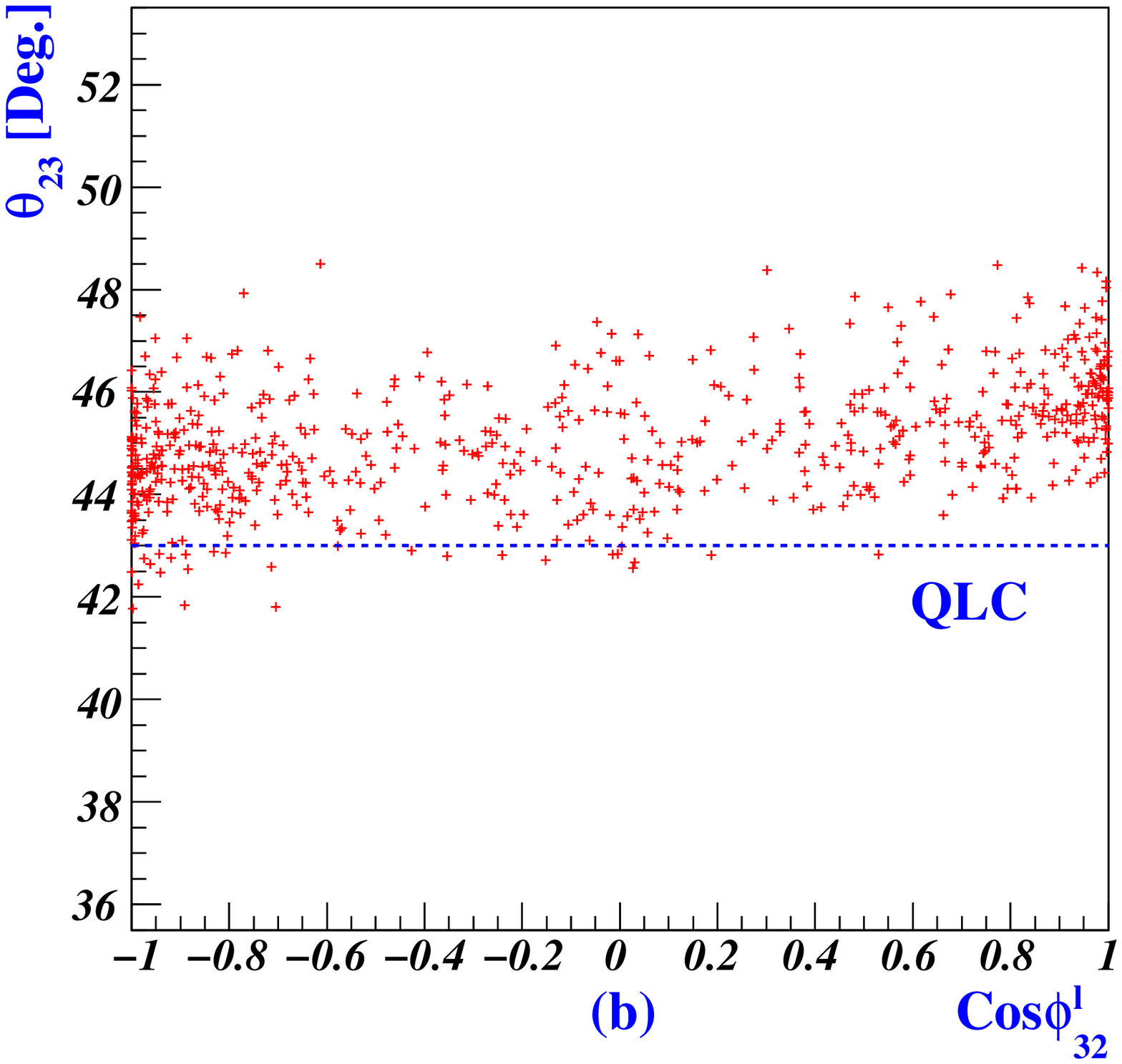,width=6.5cm,angle=0}
 \end{minipage}
 \caption{\label{FigC1}
  Same as Fig.~\ref{FigA1} except for Scenario-III. }
\end{figure}

Finally, we discuss the third scenario in which the charged leptonic mixing matrix $V^{\ell}_L$ is given
by Eq.~(\ref{LL2}). From $U_{\rm PMNS}$ in Eq.~(\ref{PMNS}), the solar neutrino mixing angle
$\theta_{12}$ can be approximated, up to order $\lambda^{3}$, as
 \begin{eqnarray}
  \sin^{2}\theta_{12} &=& \frac{1-2\lambda\cos\phi^{\ell}_{3}
   -2A_{3}\lambda^{2}\cos\phi^{\ell}_{2}+\lambda^{3}(\cos\phi^{\ell}_{3}
   +2A_{3}\cos\tilde{\phi}^{\ell}_{32})}{2+\sin2\theta\cos\varphi_{21}
   -\Xi\lambda-\lambda^{2}\Psi_{3}+\Omega_{3}\lambda^{3}}~,
  \label{angle12III}
 \end{eqnarray}
where
 \begin{eqnarray}
  \Psi_{3} &=& \Upsilon_{3}-A_{3} \Big[ \sqrt{3}\cos2\theta\sin\phi^{\ell}_{2}
   -\cos\phi^{\ell}_{2}+2\cos\phi^{\ell}_{2}\sin2\theta\cos(\varphi_{21}+\pi/3) \Big] \nonumber\\
  \Omega_{3} &=& \frac{\Xi}{2} +A_{3} \Big[ (1+2\sin2\theta\cos\varphi_{21})\cos\phi^{\ell}_{32}
   -\sqrt{3}\cos2\theta\sin\phi^{\ell}_{32} \Big] ~,
  \label{12parameterIII}
 \end{eqnarray}
with
\begin{eqnarray}
  \Upsilon_{3}= \sqrt{3}\sin2\theta\cos(\varphi_{21}+\pi/6) ~.
 \end{eqnarray}
Following the similar discussions given in Scenarios-I and -II, we can check the properties of the
above parameters in the tree level limit: if we turn off the higher dimensional operators in the
Lagrangian, that is, if $\theta \to \pm \pi/4$, $\varphi_{21} \to 0(\pi)$ and $\lambda \to 0$,
the TBM angle $\sin^{2}\theta_{12}=1/3$ is restored, as expected.
Also, in the limit of $\theta\rightarrow\pi/4$ and $\varphi_{21}\rightarrow0$ (inverted mass hierarchy
of the neutrinos), or $\theta\rightarrow-\pi/4$ and $\varphi_{21}\rightarrow\pi$ (normal mass hierarchy
of the neutrinos), the above parameters behave as $\Xi\rightarrow 0$, $\Psi_{3}\rightarrow 3/2$ and
$\Omega_{3}\rightarrow3A_{3}\cos\phi^{\ell}_{32}$.
In the left plot of Fig.~\ref{FigC1}, we show the dependence of the solar mixing angle $\theta_{12}$
on the phase $\phi^{\ell}_{3}$.  To satisfy the first QLC relation, the value of $\phi^{\ell}_{3}$
needs to be in the range of $0.174\lesssim\cos\phi^{\ell}_{3}\lesssim0.643$, which is the same range
as that obtained in Scenario-II, but larger than that found in Scenario-I.

The atmospheric mixing angle $\theta_{23}$ can be approximated as
 \begin{eqnarray}
  \sin^{2}\theta_{23}=\frac{1-\sin2\theta\cos(2\pi/3-\varphi_{21})
   -\Xi\lambda-\lambda^{2}\Upsilon_{3}+\lambda^{3}\Xi/2}{2+\sin2\theta\cos\varphi_{21}
   -\Xi\lambda-\lambda^{2}\Psi_{3}+\lambda^{3}\Omega_{3}}~.
  \label{angle23III}
 \end{eqnarray}
In the limit of $\theta \to \pm \pi/4$ and $\varphi_{21} \to 0(\pi)$, the behavior of the
parameters are found to be $\Upsilon_{3}\rightarrow3/2$, $\Xi\rightarrow0$, $\Psi_{3}\rightarrow3/2$
and $\Omega_{3}\rightarrow3A_{3}\cos\phi^{\ell}_{32}$. Then, the atmospheric mixing angle can be
approximated, up to order $\lambda^{3}$, as
 \begin{eqnarray}
  \sin^{2}\theta_{23} \approx \frac{1}{2}
   -\frac{\lambda^{2}}{2} \left( \frac{1}{2}+\lambda\cos\phi^{\ell}_{32} \right)~,
 \end{eqnarray}
where $A_{3}=1$ is used. Contrary to the cases of Scenario-I and -II, $\sin^{2}\theta_{23}$
given in Eq.~(\ref{angle23III}) is sensitive to $\cos\phi^{\ell}_{32}$ due to the absence of
$\phi^{\ell}_{1}$ [see Eq.~(\ref{LL2})]. The right plot of Fig.~\ref{FigC1} shows the behavior of
the atmospheric mixing angle $\theta_{23}$ in terms of $|\phi^{\ell}_{32}|$.
The second QLC relation is satisfied if $-1\lesssim\cos\phi^{\ell}_{32}\lesssim0.55$. Also, the lower
bound on $\theta_{23}$ is found to be $\theta_{23}\gtrsim41^{\circ}$.

\begin{figure}[t]
 \begin{minipage}[t]{6.0cm}
  \epsfig{figure=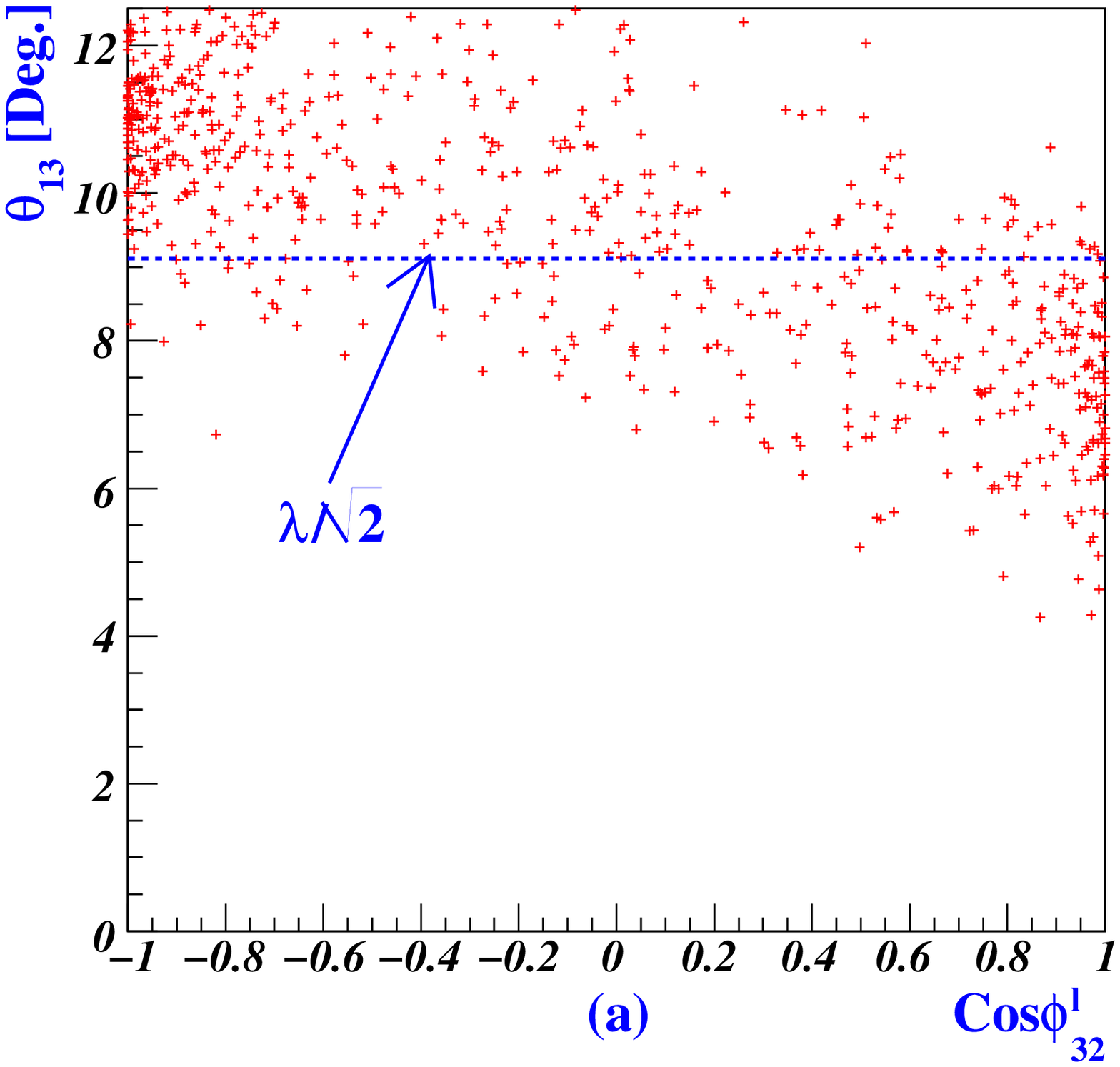,width=6.5cm,angle=0}
 \end{minipage}
 \hspace*{1.0cm}
 \begin{minipage}[t]{6.0cm}
  \epsfig{figure=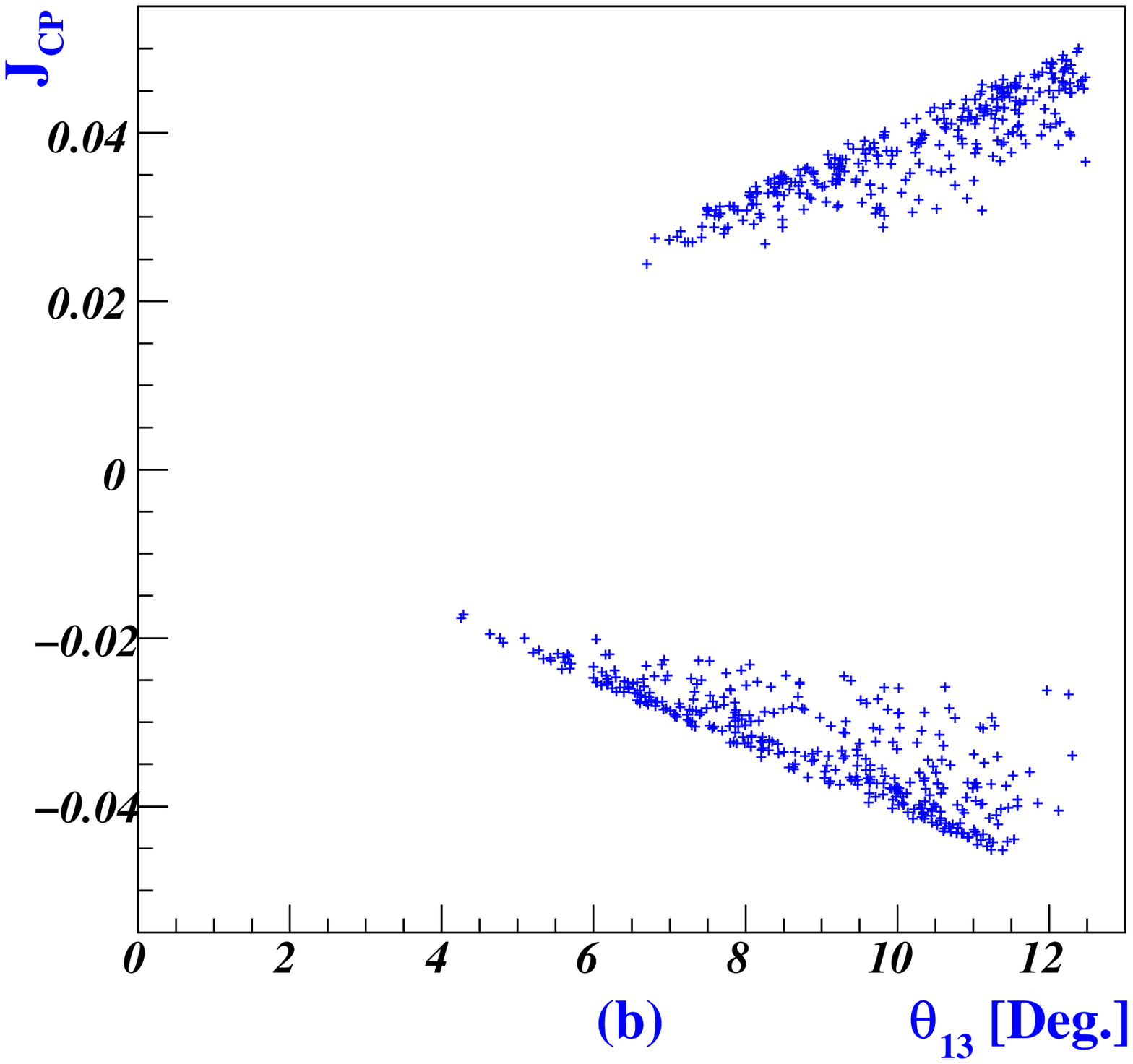,width=6.5cm,angle=0}
 \end{minipage}
 \caption{\label{FigC2}
  Same as Fig.~\ref{FigA2} except for Scenario-III. }
\end{figure}

The reactor angle $\theta_{13}$ can be expressed approximately as
 \begin{eqnarray}
  \sin\theta_{13}
  =\frac{1}{\sqrt{3}}\sqrt{1-\sin2\theta\cos\varphi_{21}
   +\Xi\lambda+\lambda^{2}\Psi_{3}-\lambda^{3}\Omega_{3}}~.
  \label{Theta13III}
 \end{eqnarray}
In the limit of $\theta \to \pm \pi/4$ and $\varphi_{21} \to 0(\pi)$, the behaviors of the
parameters $\Xi, \Psi_3, \Omega_3$ are the same as before, which leads to $\sin\theta_{13} \to
\lambda\sqrt{\frac{1}{2}-\lambda\cos\phi^{\ell}_{32}}$.
Then the reactor anlge can be rewritten as
 \begin{eqnarray}
  \sin\theta_{13}\simeq\sqrt{\frac{\lambda^{2}}{2}-\lambda^{3}\cos\phi^{\ell}_{32}
   +\frac{\varepsilon+\Xi\lambda}{3}}
 \end{eqnarray}
which indicates that the lower bound on $\sin\theta_{13}$ is obtained when $\cos\phi^{\ell}_{32} =1$.
The left plot of Fig.~\ref{FigC2} displays the dependence of $\theta_{13}$ on $\cos\phi^{\ell}_{32}$,
where the lower bound on $\theta_{13}$ is found to be $\theta_{13}\gtrsim4^{\circ}$ for
$\cos\phi^{\ell}_{32}=1$.
By comparing the left plot of Fig.~\ref{FigC2} with the right one of Fig.~\ref{FigC1}, we find that
$\theta_{13}\gtrsim5^{\circ}$ and $\cos\phi^{\ell}_{32}\lesssim0.55$ are favored to satisfy the
second QLC relation.
In the right plot of Fig.~\ref{FigC2}, we show our prediction for the Jarlskog invariant in the
leptonic sector given in Eq.~(\ref{JCP3}): $|J_{\rm CP}| \approx 0.018-0.05$.

\section{Conclusion}

The two QLC relations, given by $\theta_{12}+\theta^{q}_{12} = \pi/4$ and
$\theta_{23}+\theta^{q}_{23} = \pi/4$, may guide us to a certain symmetry between quarks and leptons.
Motivated by the QLC relations, we have invoked the discrete $A_4$ flavor symmetry in a seesaw
framework. In this work, we have presented new scenarios that can accommodate the QLC relations
and the nonzero mixing angle $\theta_{13}$ together with all the other neutrino experimental data,
including $\Delta m_{\rm sol}^2$, $\Delta m_{\rm atm}^2$, $\sin^2 \theta_{12}$ and $\sin^2 \theta_{23}$,
in a consistent way to generate the CKM matrix for the quark mixing.
Our main ingredients are the effective dimension-5 operators, invariant under
$SU(2)_{\rm L} \times U(1)_{\rm Y} \times A_{4} \times Z_{2}$ symmetry, introduced in the neutrino,
charged lepton, and quark sectors.  By generating all the necessary off-diagonal elements of the mixing
matrices, the effects of the dimension-5 operators induce a deviation of the lepton mixing matrix from
the TBM pattern and lead the quark mixing matrix to the CKM one in form.

Based on the possible interrelation between the charged lepton and quark mixing structures in our
framework, we have explicitly constructed the lepton mixing matrix to have the particular form of the
``CKM-like matrix'' (induced from the charged lepton sector) times the ``TBM pattern matrix''
(induced from the neutrino sector), which is very different from the conventional QLC scenarios
characterized by the ``bimaximal minus CKM mixing''.
We have demonstrated in detail three scenarios corresponding to three different possibilities of
constructing the charged lepton mixing matrix and pointed out that the {\it phases} of whose elements
play a key role to satisfy the two QLC relations.  For example, we have found that the value of the
solar mixing angle $\theta_{12}$ depends strongly on the particular phase $\phi^{\ell}_3$ in all the
three scenarios, while the value of the atmospheric mixing angle $\theta_{23}$ is dependent on the phase
$\phi^{\ell}_1$ [in Scenario-I and -II] or $(\phi^{\ell}_3 -\phi^{\ell}_2)$ [in Scenario-III].
Our result shows that for the reactor mixing angle $\theta_{13}$ its possible values can vary around
the center value $\sin \theta_{13} \simeq \lambda /\sqrt{2}$ (the Cabbibo angle $\lambda \simeq 0.22$)
and have the lower bound $\theta_{13} \gtrsim 3.5^{\circ}$. We have also found
that sizable leptonic CP violation
characterized by the Jarlskog invariant $|J_{\rm CP}|\sim{\cal O}(10^{-2})$
is expected. These predictions can be tested in the future experiments such as the upcoming long baseline
neutrino oscillation ones.

\acknowledgments{
This work was supported in part by the National Science Council of R.O.C. under Grants Numbers:
NSC-97-2112-M-008-002-MY3, NSC-97-2112-M-001-004-MY3 and NSC-99-2811-M-001-038. }

\appendix
\section{Higgs Potential and vacuum alignment}

\begin{figure}[h]
  \epsfig{figure=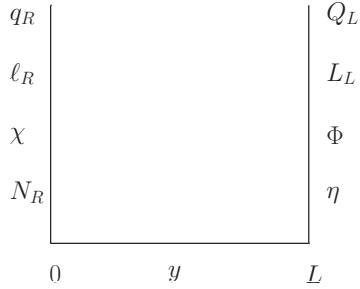,width=5cm,angle=0}
 \caption{\label{fig:exd}
  Fifth dimension and locations of scalar and fermion fields.}
\end{figure}

Since it is nontrivial to ensure that the different vacuum alignments of
$\langle\varphi^{0}\rangle=(\upsilon,\upsilon,\upsilon)$ and $\langle\chi\rangle=(\upsilon_{\chi},0,0)$
in Eq.~(\ref{subgroup}) are preserved, we shall briefly discuss these vacuum alignments.
There is a generic way to prohibit the problematic interaction terms by physically separating the
fields $\chi$ and $(\Phi,\eta)$.
Here we solve the vacuum alignment problem by extending the model with a spacial extra dimension
$y$~\cite{Altarelli:2005yp}. We assume that each field lives on the 4D brane either at $y = 0$ or at
$y = L$, as shown in Fig.~\ref{fig:exd}. The heavy neutrino masses arise from local operators at $y=0$,
while the charged fermion masses and the neutrino Yukawa interactions are realized by non-local effects
involving both branes.
A detailed explanation of this possibility is beyond the scope of this paper.

\vspace{0.5cm}

Then, the most general renormalizable scalar potential of $\Phi, \eta$ and $\chi$, invariant
under $SU(2)_{L}\times U(1)_{Y}\times A_{4}\times Z_{2}$, is given by
 \begin{eqnarray}
V_{y=L} &=& \mu^{2}_{\Phi}(\Phi^{\dag}\Phi)_{\mathbf{1}}
 +\lambda^{\Phi}_{1}(\Phi^{\dag}\Phi)_{\mathbf{1}}(\Phi^{\dag}\Phi)_{\mathbf{1}}
 +\lambda^{\Phi}_{2}(\Phi^{\dag}\Phi)_{\mathbf{1^\prime}}(\Phi^{\dag}\Phi)_{\mathbf{1^{\prime\prime}}}
 +\lambda^{\Phi}_{3}(\Phi^{\dag}\Phi)_{\mathbf{3}_{s}}(\Phi^{\dag}\Phi)_{\mathbf{3}_{s}}\nonumber\\
  &+&\lambda^{\Phi}_{4}(\Phi^{\dag}\Phi)_{\mathbf{3}_{a}}(\Phi^{\dag}\Phi)_{\mathbf{3}_{a}}
  +i\lambda^{\Phi}_{5}(\Phi^{\dag}\Phi)_{\mathbf{3}_{s}}(\Phi^{\dag}\Phi)_{\mathbf{3}_{a}}
  +\mu^{2}_{\eta}(\eta^{\dag}\eta)+\lambda^{\eta}(\eta^{\dag}\eta)^{2}\nonumber\\
  &+&\lambda^{\Phi\eta}_{1}(\Phi^{\dag}\Phi)_{\mathbf{1}}(\eta^{\dag}\eta)
  +\lambda^{\Phi\eta}_{2}(\Phi^{\dag}\eta)(\eta^{\dag}\Phi)
  +\lambda^{\Phi\eta}_{3}(\Phi^{\dag}\eta)(\Phi^{\dag}\eta)
  +\lambda^{\Phi\eta\ast}_{3}(\eta^{\dag}\Phi)(\eta^{\dag}\Phi) ~,
\label{potential1}\\
V_{y=0}  &=&\mu^{2}_{\chi}(\chi\chi)_{\mathbf{1}}
 +\lambda^{\chi}_{1}(\chi\chi)_{\mathbf{1}}(\chi\chi)_{\mathbf{1}}
  +\lambda^{\chi}_{2}(\chi\chi)_{\mathbf{1}^\prime}(\chi\chi)_{\mathbf{1}^{\prime\prime}}
  +\lambda^{\chi}_{3}(\chi\chi)_{\mathbf{3}}(\chi\chi)_{\mathbf{3}}
  +\xi^{\chi}(\chi\chi\chi)_{\mathbf{1}} ~,
\label{potential2}
\end{eqnarray}
where $\mu_{\Phi},\mu_{\eta},\mu_{\chi}$ and $\xi^{\chi}$ are of the mass dimension 1, while
$\lambda^{\Phi}_{1,...,5}$, $\lambda^{\eta}$, $\lambda^{\chi}_{1,...,3}$ and
$\lambda^{\Phi\eta}_{1,...,3}$ are all dimensionless.
From Eqs.~(\ref{potential1}) and (\ref{potential2}), it is easy to check that the vacuum stabilities
of global minima are guaranteed.

The minimum condition of the potential $V_{y=0}$ is
 \begin{eqnarray}
  \left. \frac{\partial V_{y=0}}{\partial \chi_1} \right|_{\langle \chi_1 \rangle = v_{\chi}}
  = 2v_{\chi} \Big[ \mu^2_{\chi} + 2(\lambda_1^{\chi}
    +\lambda_2^{\chi})v^2_{\chi} \Big] = 0~,
 \end{eqnarray}
and $\left. \frac{\partial V_{y=0}}{\partial \chi_{2,3}}\right|_{\langle \chi_{2,3} \rangle = 0} = 0$
are automatically satisfied.
On the other hand, the minimum conditions for the potential on the brane $y = L$ are
 \begin{eqnarray}
  \left. \frac{\partial V_{y=L}}{\partial \varphi^0_i} \right|_{\langle \varphi^0_i \rangle,
    \langle \eta \rangle}
  &=& 2v \Big[ \mu^2_{\Phi} + 2(3\lambda_1^{\Phi} + 2\lambda_3^{\Phi})v^2+(\lambda^{\Phi\eta}_{1}
    +\lambda^{\Phi\eta}_{2}+\lambda^{\Phi\eta}_{3}+\lambda^{\Phi\eta\ast}_{3})v^{2}_{\eta} \Big]
    = 0 ~,\nonumber\\
  \left. \frac{\partial V_{y=L}}{\partial \eta} \right|_{\langle \varphi^0_i \rangle,
    \langle \eta \rangle}
  &=& 2v_\eta \Big[ \mu^2_{\eta} + 2\lambda^{\eta}v^{2}_{\eta} +(\lambda^{\Phi\eta}_{1}
    +\lambda^{\Phi\eta}_{2} +\lambda^{\Phi\eta}_{3}+\lambda^{\Phi\eta\ast}_{3})v^2 \Big] = 0 ~,
 \end{eqnarray}
where $\langle \varphi^0_i \rangle = v ~ (i = 1,2,3)$ and $\langle \eta \rangle = v_{\eta}$ are used.
We obtain three independent equations for the three unknowns $v$, $v_{\eta}$ and $v_{\chi}$.
Thus the configurations needed in our scenario can be realized at tree level.
The stability of these vacuum alignments under higher order corrections is not explored in this work.


\end{document}